\documentclass[letterpaper]{article} 
\usepackage{aaai23}  
\usepackage{times}  
\usepackage{helvet}  
\usepackage{courier}  
\usepackage[hyphens]{url}  
\usepackage{graphicx} 
\usepackage{multicol}
\usepackage{tabularx}
\usepackage{natbib}  
\usepackage{caption} 
\usepackage[x11names]{xcolor}
\usepackage{multirow}
\usepackage{cellspace}
\newcommand{\done}[1]{\textcolor{black}{#1}}

\newcommand{\refine}[1]{\textcolor{black}{#1}}

\newcommand{\newtext}[1]{\textcolor{black}{#1}}
\newcommand{\camera}[1]{\textcolor{black}{#1}}

\newcommand{\specialcell}[2][c]{%
  \begin{tabular}[#1]{@{}l@{}}#2\end{tabular}}

\AtBeginDocument{%
  \providecommand\BibTeX{{%
    \normalfont B\kern-0.5em{\scshape i\kern-0.25em b}\kern-0.8em\TeX}}}

\usepackage{breqn}
\usepackage{graphicx}
\usepackage{comment}
\usepackage{enumitem}
\usepackage{subcaption}
\usepackage{booktabs}

\usepackage{newfloat}
\usepackage{listings}

\title{Characterizing Fake News Targeting Corporations}

\author{
    Ke Zhou\textsuperscript{\rm 1,2},
    Sanja \v{S}\'{c}epanovi\'{c}\textsuperscript{\rm 1},   Daniele Quercia\textsuperscript{\rm 1,3}
}
\affiliations {
    \textsuperscript{\rm 1} Nokia Bell Labs, UK\\
    \textsuperscript{\rm 2} University of Nottingham, UK\\
    \textsuperscript{\rm 3} King’s College London, UK\\
    ke.zhou@nokia-bell-labs.com
}

\setdescription{leftmargin=0pt}

\DeclareCaptionStyle{ruled}{labelfont=normalfont,labelsep=colon,strut=off} 
\usepackage{bibentry}

\begin{document}

\maketitle






\begin{abstract}
  
Misinformation proliferates in the online sphere, with evident impacts on the political and social realms, influencing democratic discourse and posing risks to public health and safety. The corporate world is also a prime target for fake news dissemination.
While recent studies have attempted to characterize corporate misinformation and its effects on companies, their findings often suffer from limitations due to qualitative or narrative approaches and a narrow focus on specific industries. To address this gap, we conducted an analysis utilizing social media quantitative methods and crowd-sourcing studies to investigate corporate misinformation across a diverse array of industries within the S\&P 500 companies.
Our study reveals that corporate misinformation encompasses topics such as products, politics, and societal issues. 
\camera{We discovered companies affected by fake news also get reputable news coverage but less social media attention, leading to heightened negativity in social media comments, diminished stock growth, and increased stress mentions among employee reviews.}
Additionally, we observe that a company is not targeted by fake news all the time, but there are particular times when a critical mass of fake news emerges.
These findings hold significant implications for regulators, business leaders, and investors, emphasizing the necessity to vigilantly monitor the escalating phenomenon of corporate misinformation.

\end{abstract}

\section{Introduction}

Emerging research in economics and political science has raised concerns about the erosion of democracy by misinformation on social media \cite{allcott2017social}.  Similarly, concerns about misinformation effects on public health are increasing \cite{naeem2021exploration,melchior2022health}. 
Yet another sphere experiencing threat due to misinformation is corporations and capital markets \cite{adriani2019fake,fox2020fake,jahng2021fake}. There are various types of corporate fake news depending on whether the news content has the intention to deceive and harm \cite{park2020understanding}. 
When fake news has a low intention to harm and deceive, the news contents might just provide inaccurate information due to honest mistakes or negligence (e.g.,~inaccurate product release leakage).
However, some fake news has the intention to harm the company but not to deceive, i.e.,~promoting factual but negative news used to inflict harm on the competing organization (e.g.,~layoffs), while others focus on both deceiving and harming, i.e.,~manipulation of information (e.g., company executive scandal) that purposefully aims to mislead and misinform the audience/customers. All those types of corporate fake news may have an impact to different extents on a company's reputation and its capital markets.
Research on corporate fake news is limited due to a lack of data, incorporating narrative \cite{adriani2019fake,camacho2020disinformation}, qualitative \cite{serazio2021other}, and survey-based approaches \cite{cheng2020influence, bronnenberg2020consumer}, along with a focus on small parts of the corporate world \cite{kogan2019fake, chung2022theory}.

This study aims to characterize and provide empirical evidence of the determinants of corporate fake news across S\&P 500 companies, using a large-scale quantitative analysis on social media data and crowd-sourced validation. The study addresses three research questions (RQs):
\begin{description}
\item \emph{(RQ1)} {How should corporate fake news articles be categorized?} 
\item \emph{(RQ2)} {Which companies are targeted by fake news?}
\item \emph{(RQ3)} {When are companies targeted by fake news?}
\end{description}

To answer those research questions, our study makes four main contributions: 
\begin{itemize}
\item By combining social media posts from Reddit, fake news links, company stock performance, and company reviews from Glassdoor, we \emph{created a unique dataset} for investigating corporate misinformation.
We \emph{developed ten metrics} to capture the characteristics of companies and the misinformation they were subject to (\S3). 
\item We \emph{conducted a user study to fully categorize the news articles} mentioning companies in the context of misinformation. We found that a two-level taxonomy best describes these  misinformation articles and its first level includes products, politics, corporate affairs, societal issues, and brand.
Fake news articles 
mentioning high-growth companies tend to be about products and societal issues, while those mentioning  limited-growth companies tend to be about politics (\S4, \emph{RQ1}).
\item We \emph{characterized the relationship between misinformation and companies}. 
\camera{We found that companies heavily targeted by fake news often receive coverage from reputable news publishers but garner less attention on social media. However, this not only leads to more negative sentiments in public comments on social media but also links to decreased stock market growth, as well as increased stress indicators among employee reviews within these targeted companies.}
\item We found a company is not targeted by fake news all the time -- there are particular
times in which a critical mass of fake news about the company emerges (\S6, \emph{RQ3}).
\end{itemize}

\section{Related Work}


Misinformation is not a new phenomenon, as seen in historical instances like propaganda during the Second World War \cite{fox2020fake}. However, today has been dubbed a 'perfect storm' for the proliferation of misinformation \cite{starbird2021online, fox2020fake, edge2021scientific}, influenced by various factors: (i) the opportunity for anyone to become a content producer, facilitated by online platforms such as blogs and social media posts; (ii) the virality of social media, particularly for disseminating fake information \cite{vosoughi2018spread}; and (iii) advancements in deep learning natural language and visual content processing techniques, enabling the creation of fake or biased content, such as deepfakes \cite{westerlund2019emergence}, that is challenging to distinguish from truthful content. Research suggests that misinformation significantly impacts democratic and political discourse \cite{pickard2019misinformation, garimella2018political}, public health and safety \cite{swire2019public, verma2022examining}, journalism \cite{waisbord2018truth}, and exacerbates social divisions \cite{cover2022marginalising}, among other effects on the social and political spheres. Another sphere in which misinformation poses a rising threat is the corporate world \cite{jahng2021fake,adriani2019fake}. As an example, COVID-19 has been falsely linked to 5G leading to the destruction of 5G towers \cite{ahmed2020covid,moshood2020covid}.

Arguably the most important effect of fake news on companies is  brand reputation \cite{jahng2021fake,di2020fake,mills2019brand,castellani2017fake}. The damage to the brand reputation is considered a risk by 98\% of professionals, and as many as 80\% of them had to deal with at least one crisis resulting from fake news  \cite{camacho2020disinformation}. Another effect of corporate fake news is monetary (e.g., through market manipulation \cite{lin2016new}, decline in sales \cite{adriani2019fake}, or loss of share value \cite{castellani2017fake}). A successful misinformation campaign for corporate giants such as Apple or Amazon could have monetary impact measured in the billions of dollars \cite{adriani2019fake}. 
It has been shown in a small-scale study \cite{xu2021corporate} that fake news perpetrators can come both from people external to the company but also from employees themselves if they want to excerpt pressure on the company. In addition, the company that responds promptly to corporate misinformation can appropriately reduce the impact on its brand reputation and stock price \cite{mills2019brand, bronnenberg2020consumer}.

While initial evidence \cite{pwc} about the effects of fake news on companies starts to mount, current research on this topic uses mostly narrative \cite{adriani2019fake,camacho2020disinformation}, qualitative \cite{serazio2021other}, and survey-based approaches \cite{cheng2020influence}, or focuses on a small part of the corporate world (e.g., financial market \cite{kogan2019fake}). Quantitative large-scale research is still nascent owing to 
the methodological challenges in quantifying both the fake news \cite{al2021methodological} and their effects on companies \cite{adriani2019fake,jahng2021fake,xu2021corporate,di2020fake}. Our current research addresses this research gap by offering insights into the fake news targeting S\&P 500 companies shared on Reddit from 2016 to 2019. 
\done{Notwithstanding the findings of a recent study \cite{eady2023exposure} indicating that exposure to misinformation has limited influence on political outcomes, it is imperative to acknowledge and discern the unique dynamics inherent in the corporate context, where exposure to misinformation might have more significant effects regarding internal or external corporate reputation, and consequently stock valuation. We believe, unlike political beliefs and attitudes that are often deeply ingrained and difficult to change, consumers, employees and investors may be more easily swayed by false information that impacts their perceptions of a company's products or services.}

\section{Data and Methods}
To examine and compare the susceptibility of S\&P 500 companies to misinformation, we collated four types of data. We collected Reddit comments mentioning those companies along with news articles (\S\ref{sec:reddit}). The credibility of these articles was evaluated utilizing established fact-checking sources (\S\ref{sec:newsarticles}). It is possible that a company's likelihood of being targeted by fake news may simply be related to its level of newsworthiness. This news coverage may stem from an external interest in the company, such as financial performance as measured by stock growth (\S\ref{sec:stocks}), or internal factors within the company. Internal company affairs  have so far been harder to characterize. To do so, we analyzed data from a reputable company review website, Glassdoor (\S\ref{sec:glassdoor}).
Finally, using these four data sources, we computed ten metrics for each company (\S\ref{sec:company_metrics}).  

\subsection{Reddit Comments with Company News}\label{sec:reddit}

\done{We chose Reddit to collect public perception of news articles related to companies. This is because: i) Reddit serves as a social news aggregation site, ideal for studying news distribution and public response, and ii) unlike typical social networks (e.g.,~Twitter), Reddit lacks friends or followers, allowing for a more indicative reflection of mainstream news production and consumption, free from social network biases like echo chambers or filter bubbles.}
We used publicly available Pushshift API \cite{baumgartner2020pushshift} to download the  Reddit comments from January 2016 to December 2019. We included all comments from all public and quarantined subreddits (e.g., r/the\_Donald). 
Since we studied S\&P 500, we chose to focus only on the users who are from the U.S.

\begin{figure}[t]
\centering
\includegraphics[width=0.35\textwidth]{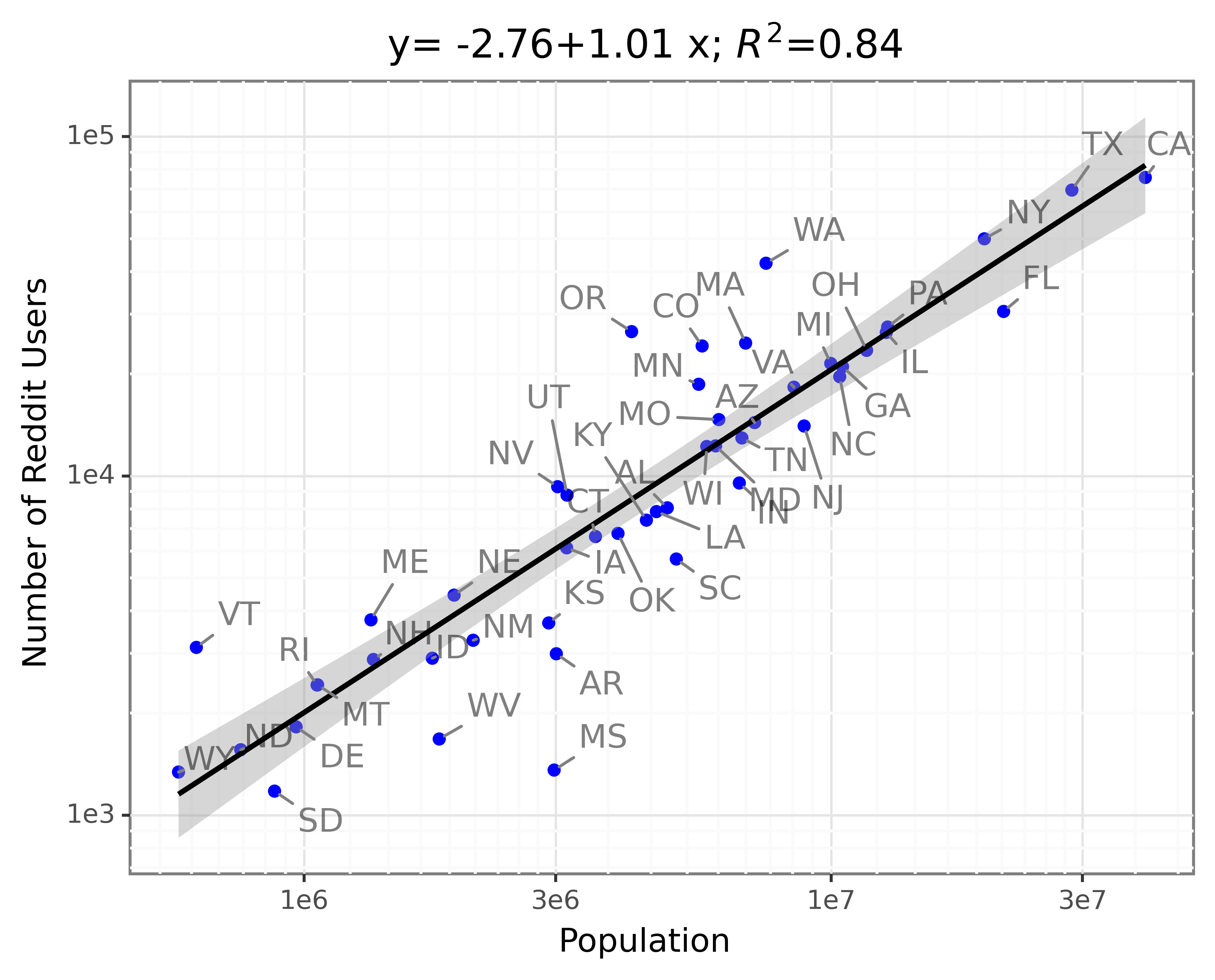}
\caption{Reddit users versus U.S. state population.}
\label{fig:users_vs_pop}
\end{figure}

Reddit does not explicitly provide user location. However, it is possible to obtain reliable estimates \cite{balsamo2019firsthand}. Starting from a list of 2,844 location-based subreddits that map to one of the U.S. cities or states (e.g., r/newyork, r/california), we listed the users with at least 5 comments in those subreddits. If they had posted in multiple states, we assigned the user to the state with the majority of comments. In this way, we placed $\sim$3M users in one of the U.S. states. The number of Reddit users per state scaled linearly with the state’s population ($\beta$ = 1.01, $R^2$=0.84) showing a good geographic representativeness (Figure \ref{fig:users_vs_pop}). 

\begin{figure}[t]
\centering
\includegraphics[width=0.4\textwidth]{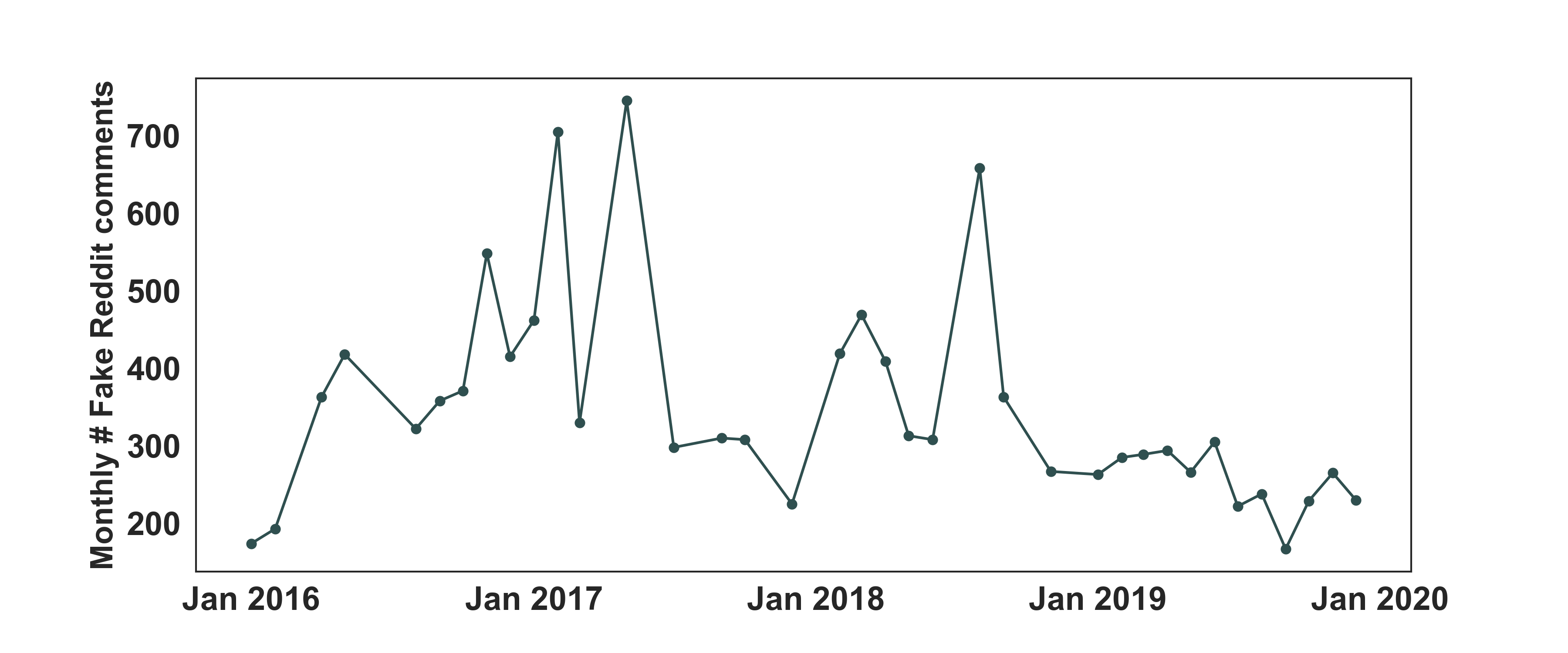}
\caption{Monthly \# of Reddit comments containing both a fake link and a company name (median: 408, standard deviation: 136). The peaks observed in early 2017 and late 2018 were mainly caused by news about Google and  Facebook. They were accused of not complying with the law that requests the removal of harmful content (Facebook) and using politically biased algorithms (Google).} 
\label{fig:data_timeline}
\end{figure}

Between 2016 and 2019, these users authored 1.4 billion comments. Of these comments, 8.2M (0.6\%) contained news links, of which 1.4M (16\%) mentioned also an S\&P 500 company (Table \ref{tab:reddit_stats}). Those comments contained 11.3M/1.3M total/unique links and had 262 (52.4\%) of the S\&P 500 companies mentioned \cite{wikisp}. Finally, we classified a Reddit comment as a fake news comment if it contained a URL to a domain that appeared in the list of fake news domains, which was defined using the credibility sources described next (\S\ref{sec:newsarticles}). In total, there were 12.8K fake Reddit comments mentioning S\&P 500 companies. We can observe the median number of Reddit comments per month discussing corporate fake news is $408$ (Figure \ref{fig:data_timeline}).

\begin{table*}[t]
\begin{center}
\tiny
\caption{Statistics about Reddit comments mentioning companies and  news links present in the comments.}
\label{tab:reddit_stats}
\begin{tabular}{ l|rrrrrr } 
 \hline
\textbf{Type of comments} & \textbf{\# comments} & \textbf{\# users}  & \textbf{\# total links} & \textbf{\# unique links} & \textbf{\# domains} & \textbf{\# companies} \\  \midrule
 comments by geolocated users in the U.S.	& 1.4B &  3M &   & &  & \\
company comments  & 11.1M	& { {795K} }& & & & 262 \\
company comments with news & 1.4M	& 322K & 11.3M & {1.3M} & 5.1K & 208 \\ 
company comments with fake news & 12.8K	& 6.3K & 145.5K & {{39.7K} }  & 316 & 114 \\\hline
\end{tabular}
\vspace{2em}
\caption{Statistics about Glassdoor reviews of S\&P 500 companies.} 
\tiny
\label{tab:glassdoor_stats}
\begin{tabular}{ llllll } 
 \hline
\textbf{\# total reviews} & \textbf{\# US-based reviews} & \textbf{\# companies}  & \textbf{avg (min-max) rating} & \textbf{avg (min-max) stess score} &  \textbf{time period} \\  \midrule
 713K & 440K &  399 &  3.37 (1.62-5) & 1.11\% (0-9.52\%) & 2008-2020 \\
\hline
\end{tabular}
\end{center}
\end{table*}

\subsection{News Credibility Sources}\label{sec:newsarticles}

We established the credibility of news at the level of the publisher (or domain). Specifically, all articles published on a domain deemed as fake or of low credibility were also considered fake or not reputable. To determine domain credibility, we employed a commonly referenced list of news domains in misinformation research \cite{bozarth2020higher}. Specifically, we labeled domains as reputable using lists compiled by Alexa \cite{alexa}, Media Bias/Fact Check \cite{mediabias}, and Vargo et al.\cite{vargo2018agenda}. This resulted in a total of $8,900$ reputable news sites. Non-reputable domains were labeled using Media Bias/Fact Check and four additional sources: Zimdars list \cite{zimdars2016my}, PolitiFact \cite{gillin2018politifact}, the Daily Dot \cite{dailydot}, and Allcott, et al~\cite{allcott2019trends}. By merging the labels from each of these five sources, we classified a domain as fake if it had a history of publishing fabricated news articles. This resulted in a total of 933 fake news domains. Upon this classification, we found that 1.4 million Reddit comments included mentions of $145.5K/39.7K$ total/unique fake news articles, as well as mentions of $114 (22.8\%)$ of the S\&P 500 companies (as outlined in Table \ref{tab:reddit_stats}).

\subsection{Company Stock Market Data}\label{sec:stocks}
We collated the S\&P 500 stock market data, including the monthly adjusted closing prices 
from 2016 until 2019 for each company, from the Yahoo Finance portal~\cite{yahoo}. \done{The stock data quantifies external perception of the company from an investor perspective.}

\subsection{Glassdoor Company Reviews}\label{sec:glassdoor}
\done{To understand internal employees' perception of their company, we acquired data regarding internal corporate affairs from 440K reviews of the S\&P 500 companies posted during the twelve years 2008-2020 on Glassdoor, 
a popular company reviewing site (Table \ref{tab:glassdoor_stats}).} In 2021, there were $50M$ monthly visitors on the platform, and $70M$ reviews about $1.3M$ companies. Current and former employees of companies describe their job experience, ranging from interviews to salaries to workplace culture. All the reviews we collected were from U.S.-based employees. We also obtained the employee ratings of each company.

\subsection{Computing Company Metrics}\label{sec:company_metrics} 

\begin{table*}[t]
    \centering
    \tiny
        \caption{List of company metrics.}
        \begin{tabular}{lSlll}
        \toprule
        {Metric} &  Formula &  Description & Distribution \\
        \midrule
        \texttt{\# Reddit comments (C, T)}  & ${\# \textrm{Reddit comments mentioning C during time period T}}$  &   public popularity & \includegraphics[height=0.5cm]{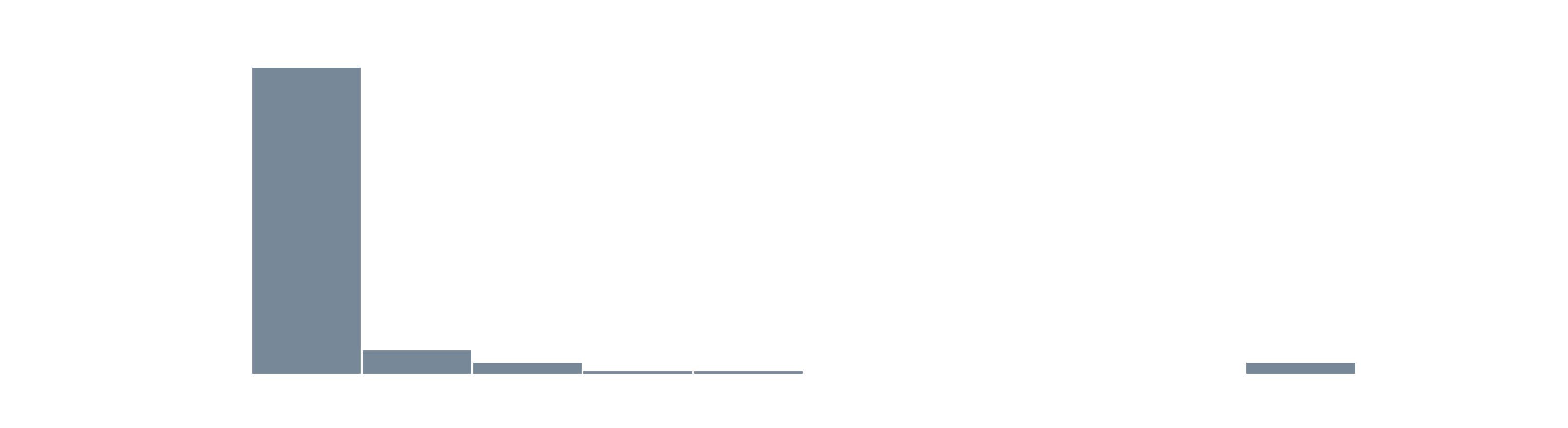} \\
        \texttt{\# Reddit comments per capita (C, T)}      &  $\frac{\# \textrm{Reddit comments mentioning C during time period T}}{\# \textrm{Glassdoor reviews of C during time period T}} $ &       corrected public popularity & \includegraphics[height=0.5cm]{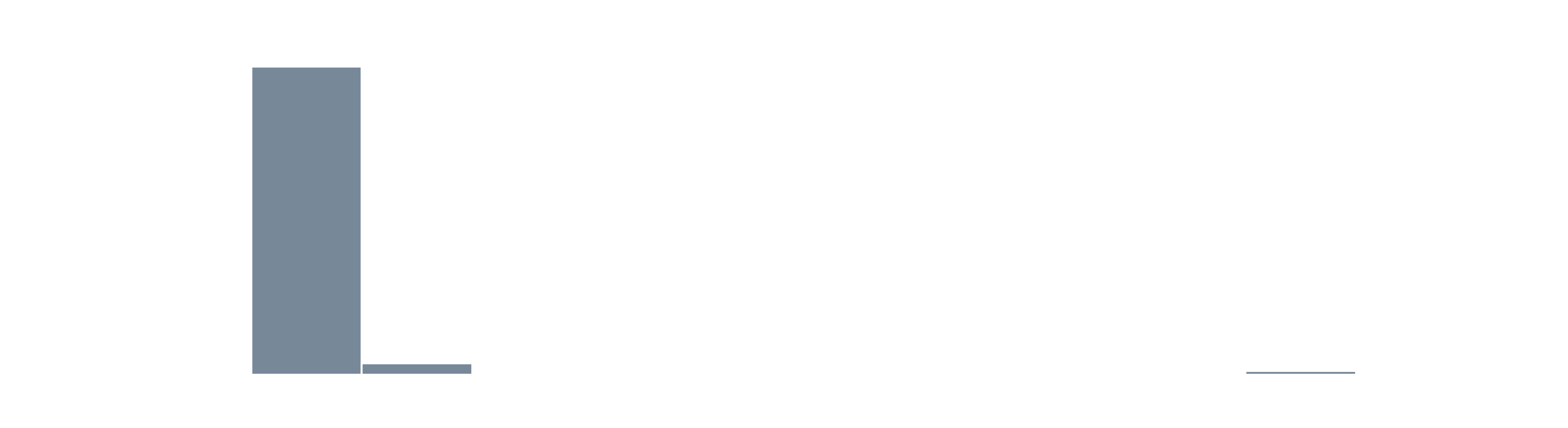} \\
        \done{\texttt{\# reputable news (C, T)}}  & \done{${\# \textrm{reputable news articles mentioning C during time period T}}$}  &   \done{newsworthiness} & \includegraphics[height=0.5cm]{figs/Reddit_comments.png} \\
        \texttt{external reputation score (C, T)}  & 1 - $\frac{\# \textrm{negative Reddit comments mentioning C in time period T}}{\# \textrm{total Reddit comments mentioning C in time period T} }$ &  \specialcell{positive/neutral mentions \\ on social media} &  \includegraphics[height=0.5cm]{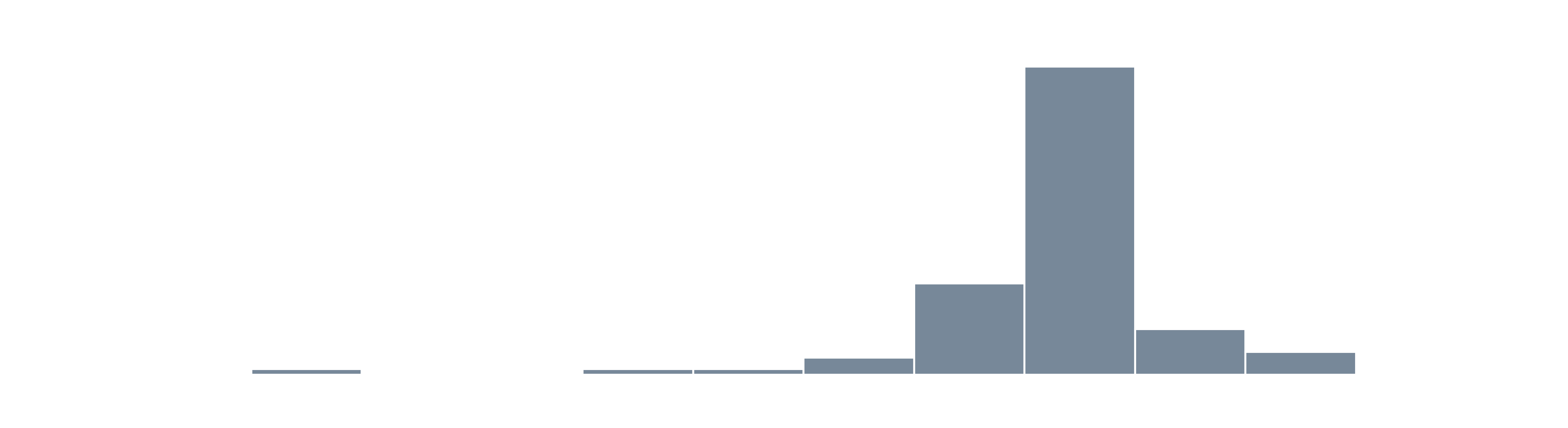}\\
        \texttt{stock valuation (C, T)} & \textrm{adjusted closing prices at end of time period $T$ for company C} &   external investor perception &  \\
        \texttt{stock growth (C)} & $ \frac{\textrm{stock valuation (C, 2019)}}{\textrm{stock valuation (C, 2014)}} $ &   external investor perception & \includegraphics[height=0.5cm]{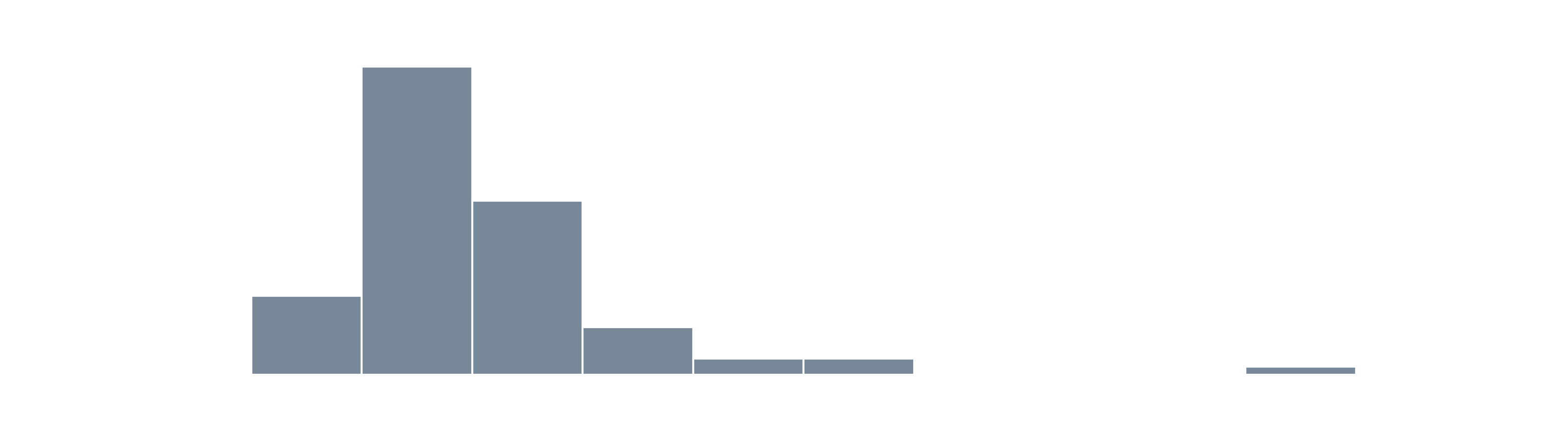} \\
        \texttt{Glassdoor rating (C, T)}        &  $\textrm{Glassdoor overall rating of company C within time period T}$  &  internal perception & \includegraphics[height=0.5cm]{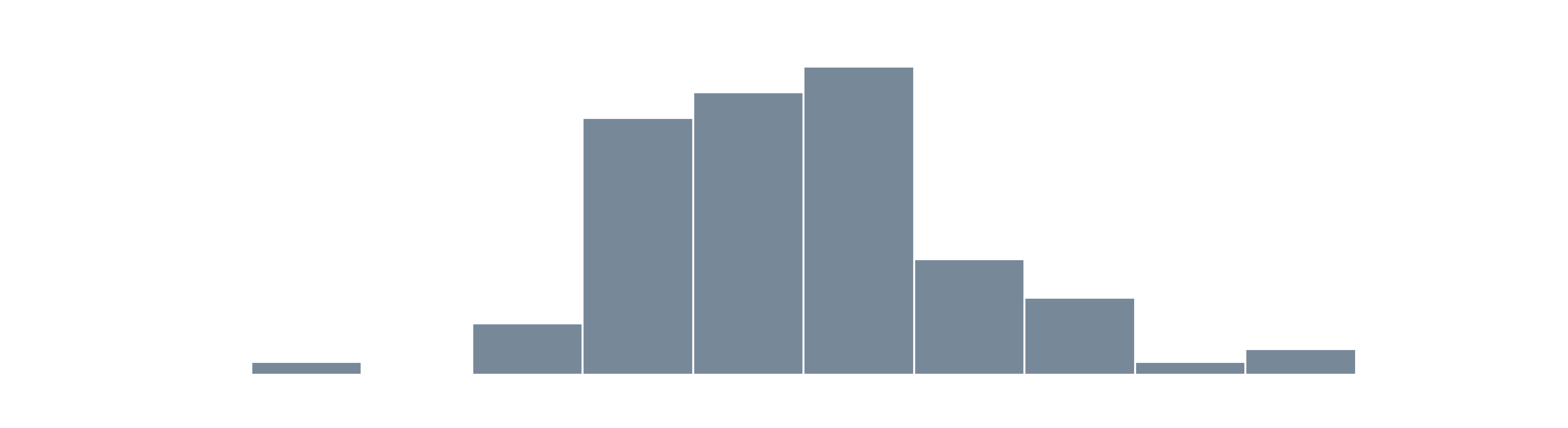} \\
        \texttt{internal employee stress score (C, T)}   &  $\frac{\# \textrm{Glassdoor reviews mentioning stress for company C within time period T}}{\# \textrm{total Glassdoor reviews for company C within time period T}}$  &  stress mentions by employees  & \includegraphics[height=0.5cm]{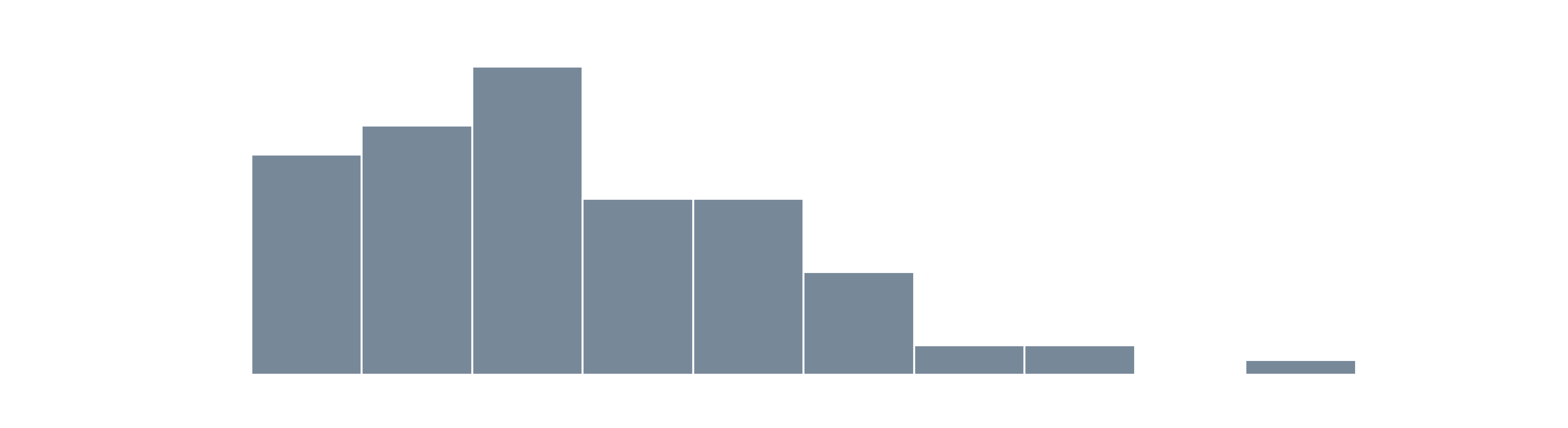} \\
        \texttt{pFake (C, T)}  & $ \frac{\# \textrm{fake Reddit comments mentioning C during time period T}}{\# \textrm{Reddit comments mentioning C during time period T}} $ & \specialcell{probability of \texttt{C} being\\targeted by fake news} & \includegraphics[height=0.5cm]{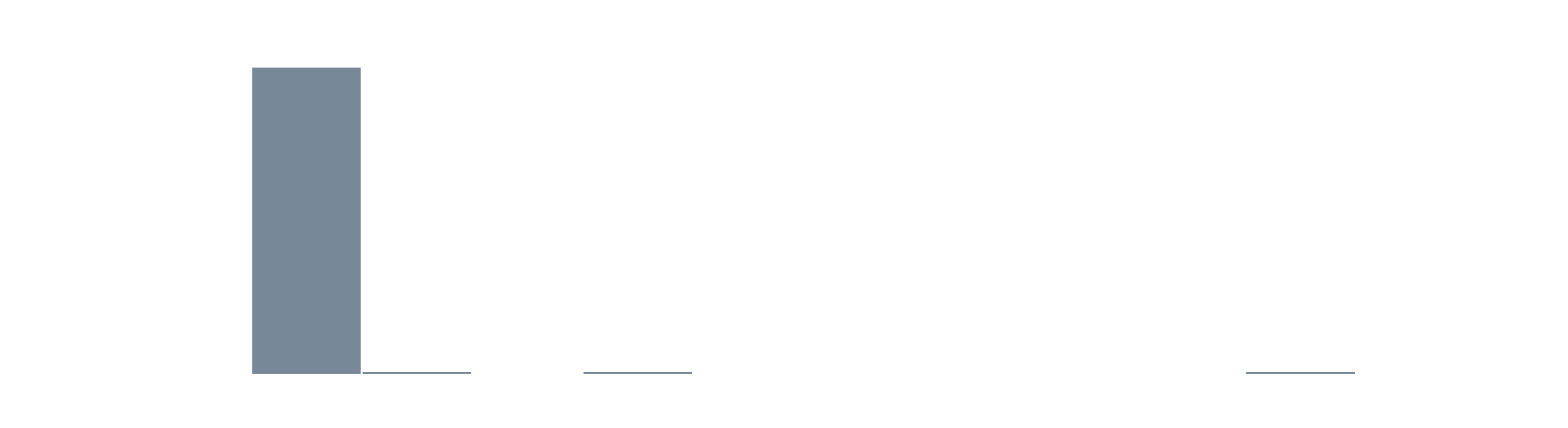} \\
        \texttt{misinformation shock (C, T)}  & $\textrm{\# months with pFake(C, T_{month})}$\textrm{$>$}$\textrm{(pFake(C, T)+2*std(pFake(C, T)))}$ & \specialcell{number of concentrated \\ misinformation attacks} & \includegraphics[height=0.5cm]{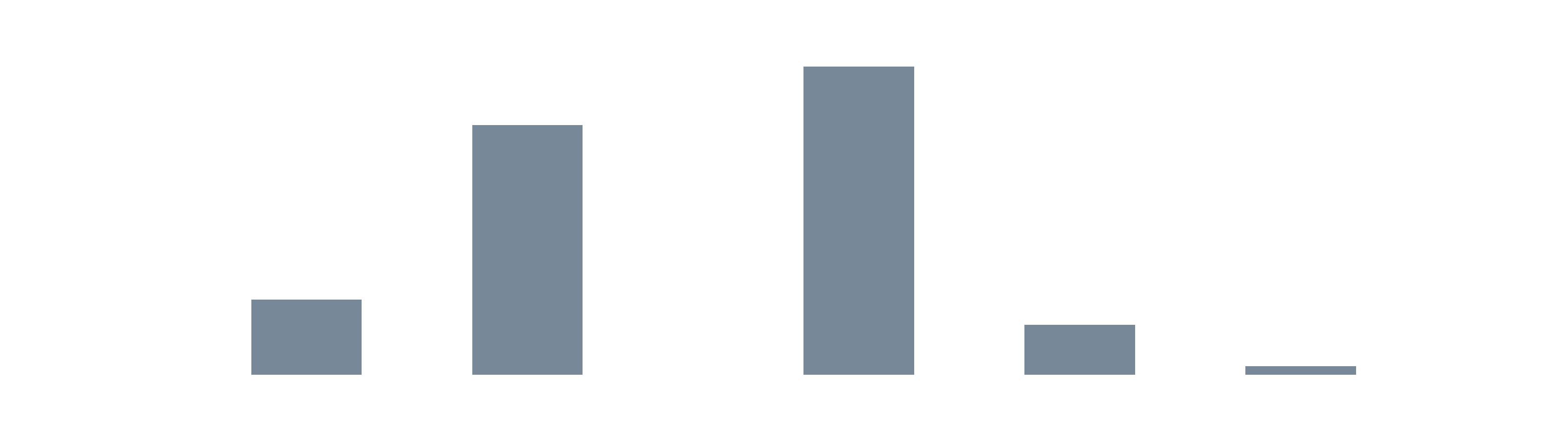} \\
        \bottomrule
        \end{tabular}

    \label{tab:res_metrics}
\end{table*}

For each company, we collated ten metrics, with formulas listed in Table \ref{tab:res_metrics}, and described below.
\begin{description}
  \item \texttt{\#Reddit comments}: we used the total number of Reddit comments discussing a company as a proxy for its popularity among the general public. To make sure that this measure does not capture simply the size of the company, we also tested a corrected metric for popularity described next.
  \item \texttt{\#Reddit comments per capita}: we calculated the total number of Reddit comments divided by the total number of Glassdoor reviews, which serves as a reasonable proxy for company's employee count.
  \item \done{\texttt{\#reputable news}: we 
calculated the total number of news articles mentioning a company in reputable news outlets, which generally captures a company's newsworthiness.}
  \item \texttt{external reputation score}: we computed \texttt{Q-score} \cite{freeman1993measuring} (one minus the fraction of the number of comments with a negative sentiment to the total number of comments mentioning the company) as a proxy for public perception of a company. To measure the sentiment of a comment, we used the Flair deep learning library \cite{akbik2019flair}, and applied it to \emph{all} the comments mentioning the company, and not only those with a news link.
  \item \texttt{Glassdoor rating}: we calculated the average rating from the company reviews on Glassdoor as a proxy for the internal perception, i.e., the company's popularity among the (former) employees. 
\item \texttt{internal employee stress score}: we calculated the percentage of Glassdoor reviews discussing stress at work as a proxy for stress mentions by employees. To extract work stress mentions from reviews, we used the MedDL entity extraction \cite{scepanovic2020extracting}, which uses contextual embeddings and a deep BiLSTM-CRF sequence labeling architecture (we used the default parameters as specified in~\cite{scepanovic2020extracting}).
The model was pre-trained and evaluated on a labeled dataset of 2K Reddit posts that have in total 4.5K medical mentions called MedRed~\cite{scepanovic2020extracting}. 
The state-of-the-art MedDL method achieved a $F1$-score of $.85$ when extracting symptoms (including stress mentions).
\item  \texttt{stock growth}: we calculated the ratio of the \texttt{stock valuation} (average adjusted closing prices) in 2019 and 2014 as a proxy for external investor perception of the company. We chose to focus on the growth of the companies between these two years for two reasons: (i) it encompasses the posting dates of the Reddit comments in our data, thus providing a relevant time frame for examination. Additionally, by selecting a stock growth starting date of 2014, two years prior to the earliest Reddit comments in our dataset (from 2016), it allows for the determination of the company's growth rate as it would have been known to Reddit users at that time; (ii) by focusing on long-term growth rather than short-term growth, we increase the robustness of the metric to potential influences of external events such as market manipulation or incidental growth/decline~\cite{forbers21}.
 \item \texttt{pFake}: we computed the probability of news articles about the company being fake as the ratio of fake to total Reddit comments with articles discussing the company.
 \item \texttt{misinformation shock}: we determined all the months with \texttt{pFake$(C, T_{month})$} two standard deviations above its mean within the entire period $T$ for that company $C$. Prior studies have shown that misinformation can be spread in a coordinated manner, where several outlets cover the same fake news in a short period of time \cite{ng2021does, muzykant2021fake}. To identify such occurrences, we identified periods of `misinformation shocks' for companies, i.e.,~these concentrated periods (months) when companies are targeted by a significant number of fake news articles. These `misinformation shocks' correspond to relatively large coordinated misinformation events.
\end{description}
We calculated each of these metrics for the entire time period $T_{all}$ between 2016 and 2019. \done{The five metrics that had heavy-tailed distributions (\texttt{\# Reddit comments}, \texttt{\# Reddit comments per capita}, \texttt{\# reputable news articles}, \texttt{stock growth} and \texttt{pFake}) were log-transformed in our analysis. }

\section{How to Categorize 
Fake News Articles?}\label{sec:categories}

Before understanding the determinants of corporate fake news, we would like to first characterize them, specifically addressing (\emph{RQ1}): \emph{How should corporate fake news articles be categorized?}
Previous studies have highlighted the importance of identifying the intentions of news outlets in the dissemination of fake news as a crucial step in combating its proliferation \cite{tandoc2019facts, kalsnes2018fake}. 

To address this question, we employed a multi-faceted approach. First, we drew upon existing literature to develop a taxonomy of fake news article categories. This was subsequently expanded through the use of manually-labeled annotations on a subset of articles, thereby identifying additional categories (\S\ref{sec:artitletype}). Second, we conducted a crowd-sourced study in which all articles were annotated with the identified categories during periods of misinformation shocks (\S\ref{sec:study}), and reported the resulting categories in \S\ref{sec:results}.

\subsection{Identifying  Categories}
\label{sec:artitletype}

Previous studies have explored the intentions and motivations behind the creation and dissemination of fake news. For instance, \citet{zhou2020survey} categorized fake news intentions as `mislead', `entertain', and `undefined', while others like \cite{kalsnes2018fake} classified motivations as `political', `financial', and `social'. Despite these endeavors, it remains unclear how these general intentions can be translated into a more nuanced and fine-grained taxonomy specifically for corporate fake news articles.

To address this gap in the literature, we conducted a qualitative study using a subset of randomly sampled $100$ corporate fake news articles. Three annotators were instructed to independently annotate the categories of these articles using the intention taxonomies from prior literature \cite{pennycook2021psychology, lazer2018science, kalsnes2018fake, zhou2020survey}. The annotators then discussed their labels to reach a final agreement on an extended taxonomy.

As shown in Table  \ref{fig:taxonomy}  (Level 1), we found that articles were primarily categorized as covering products, politics, corporate affairs, societal issues, and brands. Additionally, we added a category of ``Others" to capture any additional categories that may arise. Within each high-level level-1 category (e.g.,~product), the news articles were further categorized into more fine-grained level-2 categories (e.g.,~product issues relating to quality), as shown in Table \ref{fig:taxonomy} (Level 2). In total, there are 19 level-2 categories.

\begin{table}[t]
\caption{Companies that are targeted by fake news tend to be mentioned in articles mainly covering products, politics, corporate affairs, societal issues and brands. The top 3 and bottom 3 categories that are widely covered by news outlets are in bold and italic respectively. \newtext{For each category, the top associated moral dimension (with the highest absolute morality score) is shown in the ``Morality'' column.}}
\centering
\includegraphics[width=0.4\textwidth]{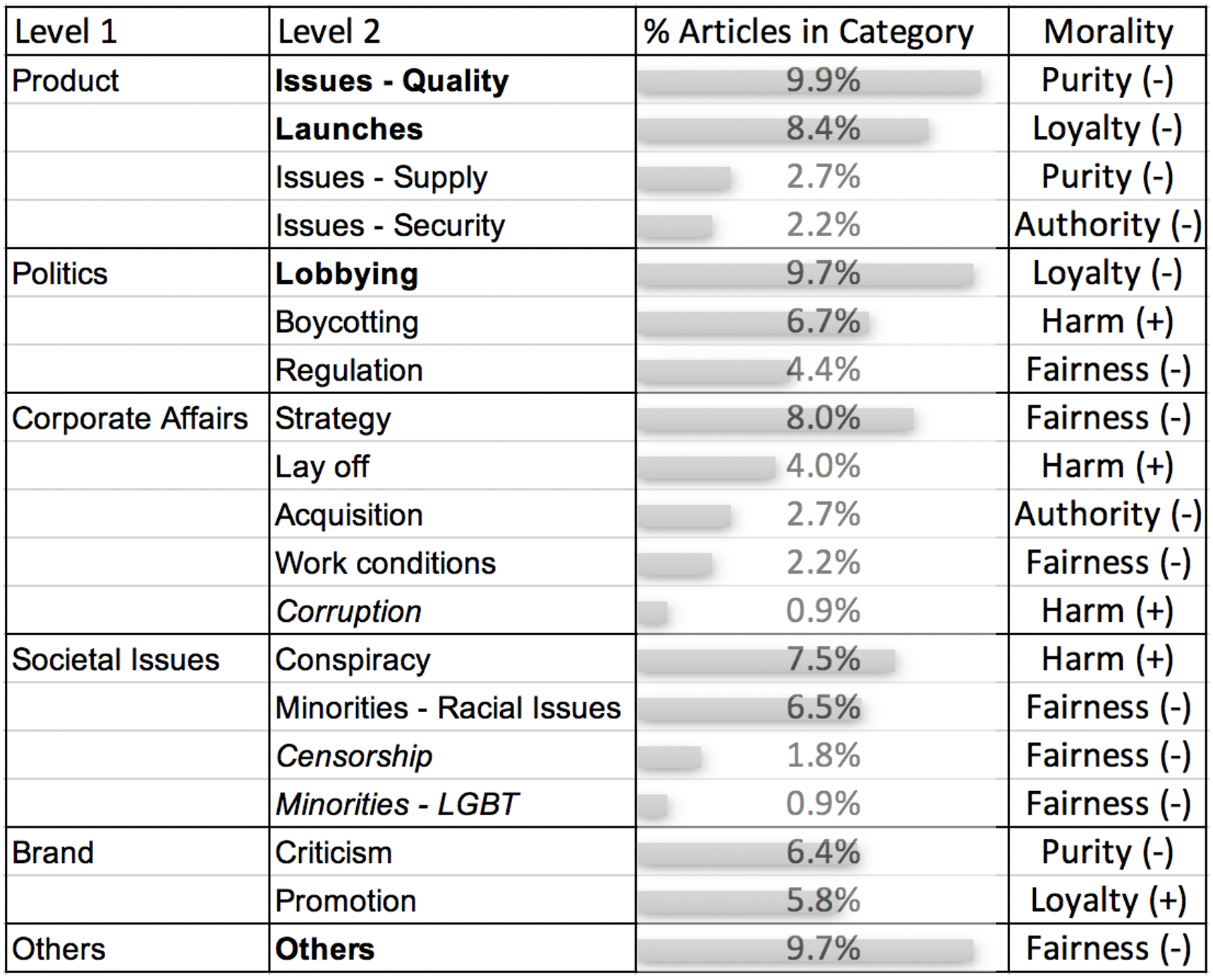}
\label{fig:taxonomy}
\end{table}

\subsection{Annotating Articles with Categories}
\label{sec:study}
We then used the manually-labelled taxonomy to label a larger subset of $N$=$1500$ salient news articles with a crowd-sourcing approach.

\mbox{ } \\ \noindent  
\textbf{Identifying articles during  misinformation shocks.}
In order to identify articles during periods of misinformation, a representative sample of fake news articles was selected. Specifically, articles that covered the companies of interest and were shared during ``misinformation shocks'' (as defined in \S3) were considered representative. To mitigate the potential for the most popular companies to dominate the results, a random selection process was employed as follows: (a)~we randomly selected a company $C$; (b)~we extracted a random set of $k=50$ articles about company $C$ shared during the months of misinformation shocks; (c)~we iterated the previous two steps until $N$ news articles were collected for annotations.
To facilitate comparison, a random sample of trustworthy news articles from the same peak periods was also obtained. This enabled the examination of both fake and trustworthy news articles.

\begin{figure}[t!]
\centering
\includegraphics[width=0.45\textwidth]{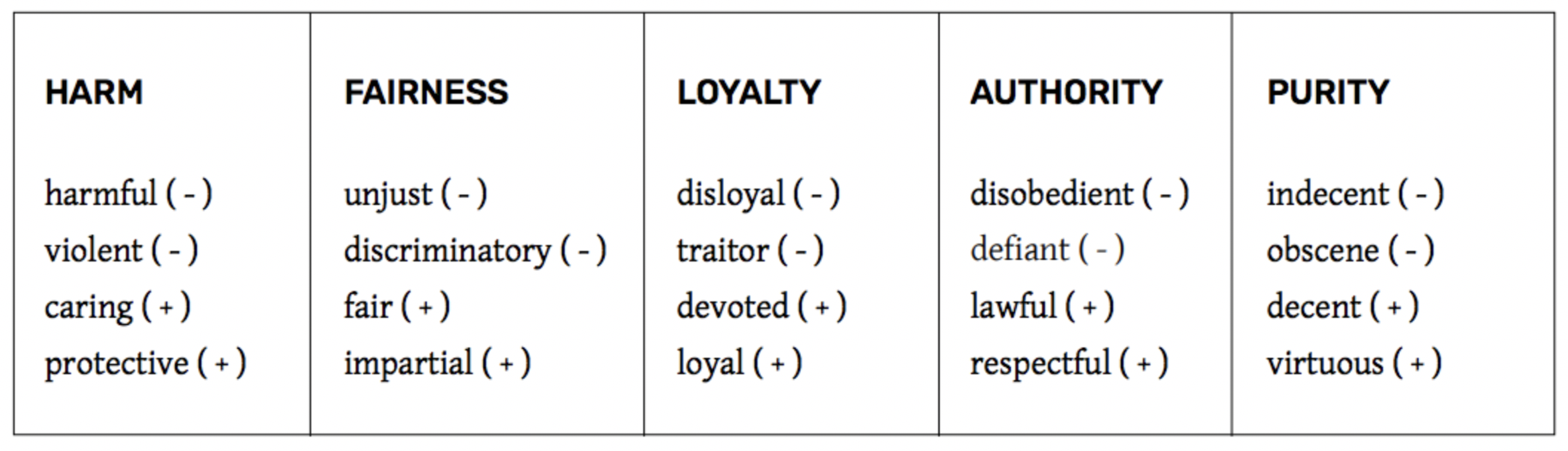}
\caption{Moral dimensions along which our participants judged a company mentioned in news articles.}
\label{fig:moral}
\end{figure}

\mbox{ } \\ 
\noindent \textbf{Crowd-sourcing data collection.}
To annotate the news articles according to the established taxonomy, a crowd-sourcing approach was employed utilizing Amazon Mechanical Turk. The data collection process consisted of the following three main steps. First, participants completed a short pre-study questionnaire in which they self-reported their age and gender. Second, participants were instructed to read the news article in detail. Third, participants were instructed to respond to three questions:

\begin{description}
\item \emph{Q1:} \emph{Which company does this article mainly cover?}
\\ (This question served to collect the company names and also served as a trap question to assess annotation quality. Participants who failed more than one trap question had their annotations removed from the analysis.) 
\item \emph{Q2:} \emph{Under which section of corporate coverage would you classify this article?}
\\ (This question collected the categories of the representative articles in Table \ref{fig:taxonomy}. If the assessors could not find any applicable category, they could add their own categorization through a free-text form under the option ``Others".) 
\item \emph{Q3:} \emph{How would you judge the company based on the article (assuming it is true)? \newtext{Please pick five words each from a different moral dimension.}} 
\\ (This question assessed how the company would be judged in the article, were it to be true, along five moral dimensions: harm, fairness, loyalty, authority, and purity \cite{graham2011mapping}. The participants were asked to pick five words each from a different moral dimension. Figure \ref{fig:moral} illustrates the positive and negative words used for those moral dimensions for the participants to pick from \cite{graham2011mapping}. We then calculated the morality score of each moral dimension by counting the corresponding number of positive words selected minus the number of negative words selected within that dimension. )
\end{description}

\done{A total of 103 participants were involved in this crowd-sourcing study, with each new article being evaluated by three participants. On average, each participant evaluated around 44 news articles. The final categorization of each article was determined using a majority vote approach. }

\subsection{
Popular News During Misinformation Shocks}
\label{sec:results}

\noindent \textbf{Categories of fake news articles.}  
The top three high-level (Level 1) categories for corporate misinformation are: \emph{Product}, \emph{Politics} and \emph{Societal} issues. Among lower-level (Level 2), the top categories are \emph{Product–Issues–Quality}, followed by \emph{Politics–Lobbying} and \emph{Product–Launches} (see Table \ref{fig:taxonomy} for a complete breakdown). Notably, the ``Others'' category accounted for only 9.7\% of fake news articles, indicating that over 90\% of fake news articles were covered by the manually established taxonomy. This partially validates the effectiveness of the taxonomy in categorizing corporate fake news articles.

\newtext{
In Table \ref{fig:taxonomy}, under the `Morality'' column, we present the top moral dimension with the highest absolute morality score for each fake news category. For instance, in news articles that cover \emph{Product-Issues-Quality,} people tend to perceive the discussed companies as morally wrong (i.e.,~negative \emph{purity}, frequently judging them as ``indecent'' and ``obscene''.).
Overall, by tracking morality scores for all fake news articles, we found that companies that are targeted by fake news tend to be judged as being unfair (i.e.,~\emph{unjust} and \emph{discriminatory} in \texttt{fairness}), morally wrong (i.e.,~\emph{indecent} and \emph{obscene} in \texttt{purity}), and harmful (i.e.,~\emph{harmful} and \emph{violent} in \texttt{harm}). We also compared the morality scores across different news categories. It is not surprising that, for news articles that cover \emph{Societal issues,} people perceive the mentioned companies as the most morally unacceptable, especially in terms of being unfair and causing harm to society. In comparison, the category of news articles \emph{Societal Issues-Minorities-Racial Issues} reflects the most morally wrong companies that people are disgusted with.
This result highlights the impact of fake news on people's moral perception of a company, which may not align with reality.}

\mbox{ } \\
\noindent \textbf{How categories of fake news articles differ from  trustworthy ones.} 
To measure the extent to which each category $i$ is associated with fake news as opposed to trustworthy news,  we calculated the \emph{misinformation association $MA(i)$} for each article category $i$: $MA(i) = \frac{P(i|fake)}{P(i|trustworthy)}$.
We found that the top 3 categories associated with fake news outlets are \emph{Product-Issues-Quality}, \emph{Politics-Lobbying}, and \emph{Societal Issues-Conspiracy}, while the top 3 associated with trustworthy news outlets are \emph{Brand promotion}, \emph{Corporate affairs-acquisition}, and \emph{Politics-regulation}. 
This demonstrates that certain categories of news articles, such as product-issues-quality, were more likely to be covered by \emph{fake} news outlets, compared to those \emph{trustworthy} ones.

\section{Which Companies Are Targeted?}

To answer this question, we identified which companies tend to be targeted, and which factors determined that targeting.

\begin{table*}[]
    \centering
    \tiny
    \caption{
    Differences for company metrics between companies that are targeted by fake news and those that are not. For each of the metrics,  we get a set of values, each corresponding to a company. Mean, median and standard deviation, and difference between targeted and non-targeted companies
    with t-stats are reported. The $p$-values for the $t$-stats significance levels are: $.1 (*)$, $.05 (**)$, and $0.01 (***) $. The statistically significant metrics are in bold. All the metrics are calculated within the time period $T_{all}$ (between 2016 and 2019). Compared to non-targeted companies, targeted ones attract more comments on Reddit, have a slightly worse \texttt{external reputation}, and \textcolor{black}{significantly more \texttt{reputable news} about them shared}.} 
    \begin{tabular}{lrrr|rrr|cl}
\toprule
& \multicolumn{3}{c}{targeted companies}&\multicolumn{3}{c}{non-targeted companies}&\\
{} &  mean &  median &  std &  mean &  median &  std &  mean diff &  $t$-stat \\
\midrule
Glassdoor rating $(C, T_{all})$      &           3.46 &             3.43 &          0.35 &              3.39 &                3.40 &             0.33 &        0.07 &     1.49 \\
internal employee stress score $(C, T_{all})$           &           0.01 &             0.01 &          0.00 &              0.01 &                0.01 &             0.01 &       -0.00 &    -0.97 \\
stock growth $(C)$   &           0.39 &             0.51 &          0.93 &              0.51 &                0.53 &             0.56 &       -0.12 &    -1.16 \\
$log$(\textbf{\# Reddit comments $(C, T_{all})$})      &           9.40 &             9.31 &          2.10 &              5.28 &                5.52 &             2.33 &        \textbf{4.12} &    \textbf{$14.22^{***}$} \\
$log$(\textbf{\# Reddit comments per capita $(C, T_{all})$})    &           1.99 &             1.99 &          2.02 &             -1.04 &               -0.78 &             2.11 &        \textbf{3.03} &    \textbf{$11.25^{***}$} \\
\textbf{external reputation score $(C, T_{all})$}              &           0.77 &             0.77 &          0.01 &              0.78 &                0.77 &             0.08 &       \textbf{-0.02} &     \textbf{$-2.13^{**}$} \\
\textcolor{black} {$log$({\textbf{\# reputable news $(C, T_{all})$}}) }         &           7.40 &             7.40 &          1.90 &              3.02 &                3.07 &             1.87 &        \textbf{4.39} &    17.82*** \\
\bottomrule
\end{tabular}
    \label{tab:res_target_vs_non_target}
\end{table*}

\subsection{Companies and Fake News Outlets}
Out of the 262 S\&P 500 companies discussed on Reddit during the entire period of study $T_{all}$, 114 were targeted by fake news, and 148 were not. What distinguishes those two types of companies? In Table \ref{tab:res_target_vs_non_target}, we show the statistics of our metrics' distributions across the two sets of companies. The two sets of companies statistically differ in their public popularity (\texttt{\# Reddit comments}, \texttt{\# Reddit comments per capita} and \texttt{\# reputable news}) and their \texttt{external reputation score}. In that, companies targeted by fake news are more popular and also discussed in more negative tone. 

\begin{figure}[t]
\label{fig:company-outlet}
\centering
\includegraphics[width=0.45\textwidth]{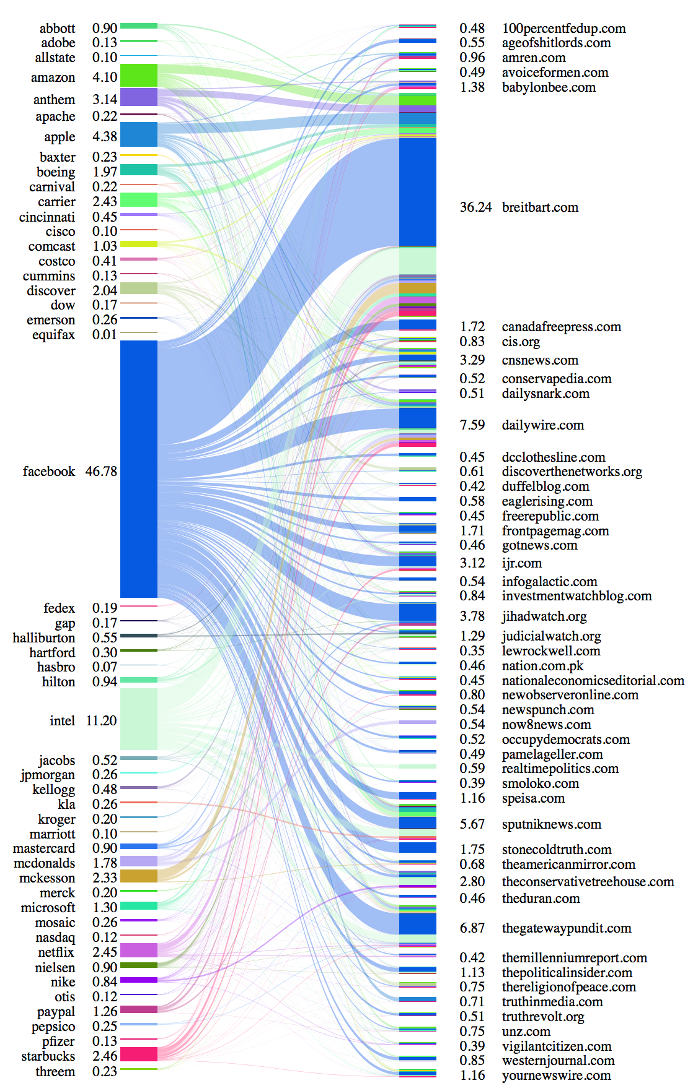}
\caption{Bipartite graph connecting companies (left hand side) with the fake news outlets (right hand side) mentioning them. A company that is likely to be target by fake news tends to be a tech company such as Facebook and Apple. Tech companies tend to be targeted by multiple outlets, while communication ones (e.g., Google) tend to be targeted by specific outlets such as \url{stonecoldtruth.com}.}
\end{figure}

To investigate the specific relationship between companies and fake news outlets targeting them, we generated a bipartite graph of company-outlet interactions in which a fake news outlet and a company are connected to each other if the company is targeted by the outlet. 
The bipartite graph for the top 50 companies and fake news outlets are shown in Figure \ref{fig:company-outlet}. We found that: (1) Internet based companies, such as Facebook, Amazon, Microsoft and Apple, attract more fake news.
(2) Certain news outlets cover many companies (e.g., \url{breitbart.com} and \url{dailywire.com}), whereas other outlets focus on reporting a particular sector (e.g., \url{stonecoldtruth.com} focuses on communication service companies).

\subsection{Factors Associated With the Targeting}

For the 126 companies targeted at least once by fake news, we calculated the probability that news discussing them are fake (\texttt{pFake $(C, T_{all})$}) and asked which company characteristics influence this probability. To that end, we ran a linear regression with \texttt{pFake $(C, T_{all})$} as the dependent variable and the company metrics in Table \ref{tab:res_metrics} as independent variables. \newtext{We confirmed that there is no multicollinearity among our independent variables using Variance Inflation Factors (VIF) \cite{miles2014tolerance}. VIF values above 5 would suggest multicollinearity issues, while the maximal VIF value we found was \done{1.44}, well below that threshold (5).}
Specifically, we ran three linear regression models, with the following sets of independent variables: 
\begin{itemize}
\footnotesize
    \item[]  \textbf{M1}: \texttt{log(\# Reddit comments per capita)},  \done{\texttt{log(\# reputable news)}}, \texttt{Glassdoor rating}  and \texttt{stock growth} 
    \item[] \textbf{M2}: \textbf{M1} + \texttt{external reputation score}
    \item[] \textbf{M3}: \textbf{M2} + \texttt{internal employee stress score}
\end{itemize}



The prediction results 
demonstrated that, despite Model \textbf{M1} being the base model, its three predictors explain around 40\% of the variance in \texttt{pFake} ($Adj. R^2$=.40). Adding the \texttt{external reputation score} in \textbf{M2} increases the explanatory power to $Adj. R^2$=.42, and also adding the \texttt{internal employee stress score} in \textbf{M3} leads to an $Adj. R^2$=.43.

Taking the best linear regression model (\textbf{M3}), we inspected its $\beta$ coefficients (Table \ref{tab:res_lr_coef}). The coefficient $p$-values are significant for company popularity (\texttt{\# Reddit comments per capita}), newsworthiness (\texttt{\#  reputable news}), and external valuations (\texttt{stock growth}). Companies that enjoy higher popularity among the general public (on Reddit) and reputable news publishers, yet possess lower stock valuations, tend to be more associated with fake news.

While the coefficients for \texttt{external reputation} and \texttt{internal employee stress score} appear insignificant, there are increasing scores from $M1$ to $M2$ (when we added \texttt{external reputation}), and from $M2$ to $M3$ (when we added \texttt{internal employee stress score}). When considering these scores' correlation with the number of reputable news sources (\texttt{\# reputable news}) and factoring in the results from the subsequent cross-lagged analyses reported in (\S6.2), it suggests their association with fake news.

To sum up, companies that suffer severely from fake news are considered newsworthy, garnering more attention from reputable news publishers, yet are perceived negatively both internally and externally.
Internally, employees are more likely to comment on a stressful work environment. Externally, investors do not highly value these companies and the general public on social media talks less about them; and if they do mention them, they do so using negative sentiment.

\begin{table}[]
    \centering
    \tiny
        \caption{Coefficients and statistics for M3 predicting \texttt{pFake $(C, T_{all})$} from our company metrics. All metrics are calculated within the time period of $T_{all}$ (between 2016 and 2019). A company that is likely to be targeted by fake news {\color{black} is the subject of more reputable news as well (i.e., it tends to be more newsworthy)}, and is not doing well in the stock market (lower stock growth). The significant coefficients with $p$-values at significance level above $.1$ are shown in bold.}

\begin{tabular}{lrrr}
\toprule
{} &   coef &  std. error &  p-value \\
Var\_name                &        &             &          \\
\midrule
\midrule
Intercept               & \textbf{-4.058} &       0.865 &    0.000 \\
\texttt{log(\# Reddit comments per capita  $(C, T_{all}))$}      & \textbf{-4.603} &       0.702 &    0.000 \\
\textcolor{black}  {\texttt{log(\# reputable news $(C, T_{all})$)}}              &  \textbf{2.802} &       0.826 &    0.001 \\
 \texttt{stock growth $(C)$}  & \textbf{-1.634} &       0.850 &    0.059 \\
 \texttt{external reputation score  $(C, T_{all})$}            & -1.480 &       0.961 &    0.128 \\
\texttt{internal employee stress score  $(C, T_{all})$}          &  1.116 &       0.722 &    0.127 \\
\texttt{Glassdoor rating  $(C, T_{all})$}       & -0.683 &       0.771 &    0.378 \\
\bottomrule
\end{tabular}
    \label{tab:res_lr_coef}
\end{table}


\begin{table}[t]
\centering
\tiny
\caption{The $Adj. R^2$ and beta coefficients of the linear regression predicting $p(article\_category | fake)$ for three categories of news articles. Companies targeted by fake news attract more comments on Reddit. However, those that are mentioned in conjunction with product and societal issues do well in the stock market as they are tech companies such as Apple, while those that are mentioned in conjunction with politics tend to do worse among their employees  (in terms of their company review ratings) and typically are pharmaceutical companies such as Pfizer. Significant values are marked with * based on their significance levels $p<0.05$.} 
     \begin{tabular}{l|c|c|c}
        \toprule
          {\bf Feature}   & \emph{Product} & \emph{Politics} &  \emph{Societal Issues} \\  \hline
        Intercept & -1.22*  & -2.14*  & -1.87*  \\ 
        \# Reddit comments per capita $(C, T_{all})$ & 2.41* & -0.73 & 1.63* \\ 
        external reputation score $(C, T_{all})$ & -0.35 & 0.72 & -0.55  \\ 
        Glassdoor rating $(C, T_{all})$ & -0.89 & -1.24* & -0.76  \\ 
        internal employee stress score  $(C, T_{all})$ & 0.12 & 0.24 & -0.10 \\  
        stock growth $(C)$ & 1.45* & 0.54  & 1.02*  \\\hline
        Adjusted $R^2$ & 0.27 & 0.20 & 0.16 \\ 
                \bottomrule
    \end{tabular}
    \label{tab:models}
\end{table}

\subsection{Factors that Influence the Most Popular Categories of Fake News Articles}

To understand the determinants that impact only specific categories of fake news articles (Table \ref{fig:taxonomy}), 
we ran three linear regressions having the probability of a fake news article belonging to each of the top 3 categories (\emph{Product}, \emph{Politics}, and \emph{Societal Issues}, as defined in \S\ref{sec:results}) as a dependent variable, and the five company metrics of \texttt{\# Reddit comments per capita}, \texttt{internal employee stress score}, \texttt{external reputation score}, \texttt{stock growth}, and \texttt{Glassdoor rating} as independent variables.
Results in Table \ref{tab:models} show 
companies that are mentioned in conjunction with the \emph{Product} and \emph{Societal Issues} attract more comments on Reddit and do well in the stock market as they tend to be tech companies such as Apple. This indicates when covering high-growth companies, fake news outlets tend to cover popular and controversial topics. However, companies linked to fake news in the realm of \emph{Politics} tend to receive lower ratings from their employees.  -- these typically are pharmaceutical companies such as Pfizer.

\begin{figure*}[t]
\centering
\includegraphics[width=0.98\textwidth]{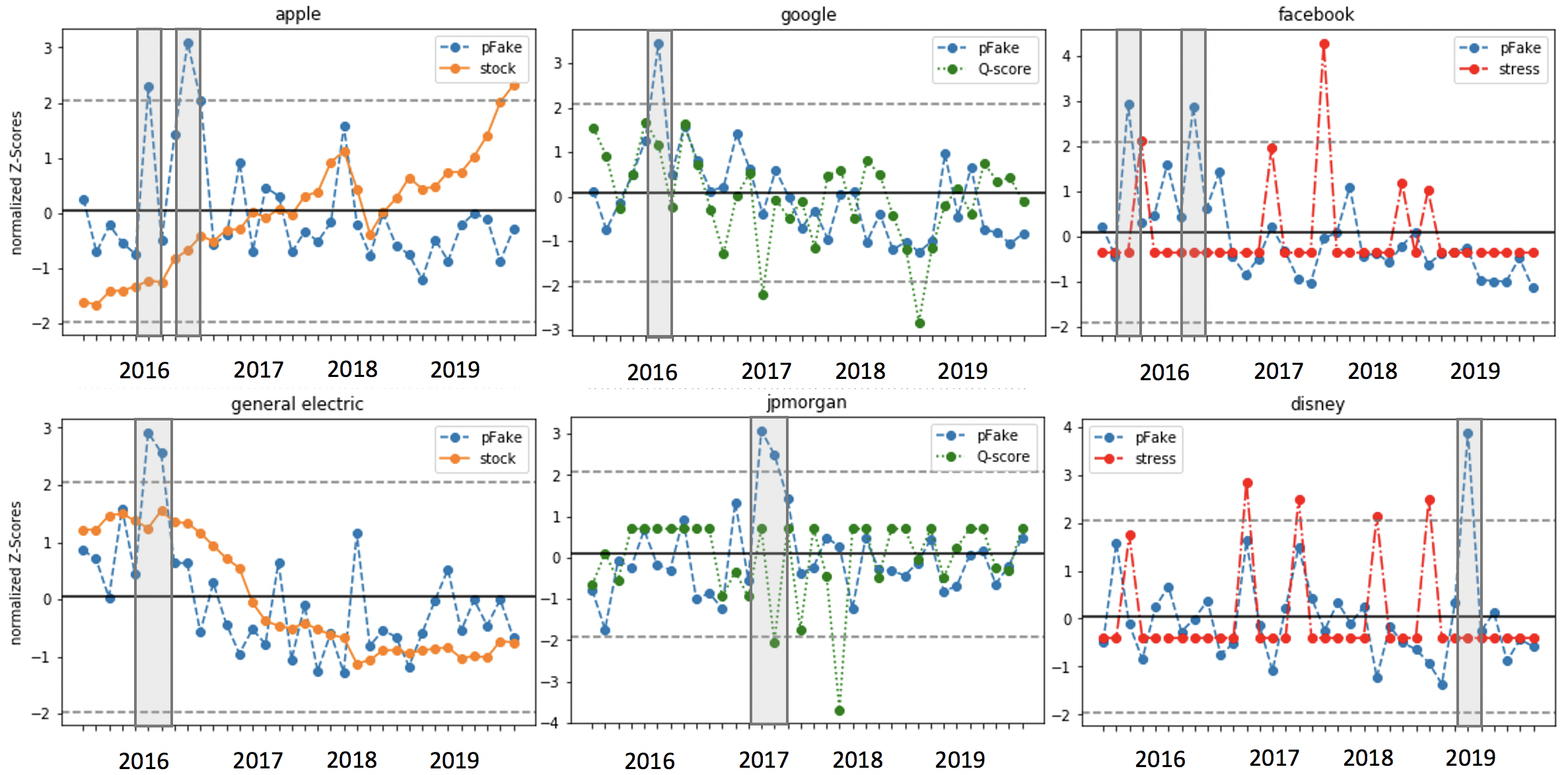}
\caption{
\refine{Fake news shocks (peaks of $pFake(C, T_{month})$ over time, shown with the gray bars) for a company C that: is a Tech company (Apple and Google); is a Communication company (Facebook); has not done well in the stock market (General Electric); attracts considerable negative mentions on social media (JPMorgan); and attracts considerable stress mentions by its employees (Disney).}}
\label{fig:shock} 
\end{figure*}

\section{When Companies Are Targeted?}

Our next research question is to understand \emph{when} companies are targeted by fake news (\emph{RQ3}). We are particularly interested in finding out whether a company is targeted by fake news all the time, or within those concentrated periods, i.e.,~misinformation shocks (as we defined in \S\ref{sec:company_metrics}).
%

Specifically, we identified misinformation peaks for company C 
by identifying those months that have 
\texttt{$pFake(C,T_{month})$}  2 standard deviations above the mean of \texttt{$pFake(C,T_{all})$} within $T_{all}$. In Figure \ref{fig:shock}, we showed several examples of \texttt{misinformation shocks} plus its most aligned metric among the three: \texttt{stock valuation $(C, T_{month})$}, \texttt{external reputation score $(C, T_{month})$} and \texttt{internal employee stress score$(C, T_{month})$}. The companies that are shown  are those for which, based on visual inspection, misinformation shocks were  aligned the most with at least one of the three metrics. We should stress that such an alignment is a simple association and does not speak to any causal claim though.  All the metrics are normalized using $z$-score throughout the entire periods of study (2016-2019).

\subsection{Qualitative Case Studies}

\begin{description}
\item \emph{Apple}. 
The company experienced two distinct instances of misinformation, or ``misinformation shocks", in late 2016 and early 2017. The majority of the articles during these periods focused on three main topics: the FBI's alleged request for Apple to add ``back-doors" in their phones; false claims of a potential recall of their phones due to alleged issues with them catching fire; and a ransomware attempt in which hackers threatened to wipe out 300 million iPhone accounts in early 2017, however, there was no evidence that the company was affected by this attempt. After the second misinformation shock, the company's \texttt{stock valuation} decreased by 0.2 standard deviations.

\item \emph{Google}. The company experienced one instance of misinformation, or ``misinformation shock'', in late 2016. The articles during this period focused on a single claim that the company's ranking algorithm downgraded any news stories that would harm the then presidential nominee of the Democratic party, Hillary Clinton. This claim was unproven and categorically rejected by the company. As a result of this misinformation shock, the company's \emph{external reputation score} decreased by 1.4 standard deviations.

\item \emph{Facebook}. The company experienced two instances of ``misinformation shocks''. The first occurred in early 2016 and centered around the claim that the company was not complying with European laws requiring the removal of hate speech on its platform. One month after this shock, the company's \texttt{internal stress score} increased by 2.3 standard deviations. The second shock occurred in early 2017 and involved two separate events related to discrimination. The first event was an allegation that the platform had an internal document advising on how to censor news of crimes against Black people. The company acknowledged this document was no longer in use. The second event was an allegation that the company's senior engineers blocked diversity hires. The company declined to comment on this claim but stated that they ``care deeply about diversity''.

\item \emph{General Electric}. 
The company experienced a single instance of significant ``misinformation shock'' that lasted for two months in late 2016. The majority of the articles during this period focused on a rumor that the company intended to sell its \$3 billion Industrial Solutions business. After the circulation of this rumor, the company's \texttt{stock valuation} decreased by 0.2 standard deviations. It was later revealed that the rumor was true, as the sale of the business was initiated in late 2017 and completed in mid-2018.

\item \emph{JP Morgan}. The company experienced a single instance of significant ``misinformation shock'' that lasted for two months in late 2017. The majority of the articles during this period focused on a scandal in which the company, JP Morgan, was alleged to have transferred \$875 million into a former Nigerian oil minister's account, and was subsequently sued by the Nigerian government. After this, the company's \texttt{external reputation score} decreased by 2.7 standard deviations. However, in 2022, a judge at the British High Court ruled in favor of JP Morgan.

\item \emph{Disney}. The company experienced a single instance of significant ``misinformation shock'' in mid-2019. The majority of the articles during this period focused on allegations of child pornography involving former or current employees. In one instance, a former employee was arrested and charged with ``transferring obscene materials to a minor and attempting to entice a minor''.\footnote{\url{https://www.nbcnews.com/news/us-news/disney-world-employee-tried-lure-8-year-old-girl-sexual-n1009466}} 
This shock did not have a significant impact on the company's \emph{internal stress score}, unlike other minor instances of misinformation that had occurred in the past, such as slow progress in building Disney's Pandora land in 2017 and allegations of CIA aid in the company's land acquisition in Florida in 2018, which did.

\end{description}

These are just examples of a general trend that we observed: a company is not targeted by fake news all the time, but there are particular times in which a critical mass of fake news emerges, and that might impact either the company's \emph{external reputation}, \emph{internal employee stress} or even \emph{stock valuation}.

\subsection{\done{Cross-lagged Analysis}}
To further examine how fake news impact company's external reputation, internal employee stress and stock valuation, we conducted cross-lagged analysis. The goal is to determine causal relationships by examining the connection between two variables at different points in time, while taking into account the influence of earlier measurements of the same variables.
This type of analysis helps eliminate the possibility that the observed relationship between the variables is caused by a third or confounding variable. In our study, we conducted cross-lagged analysis on all time pairs ($t_k$, $t_l$) and selected the pair and the model based on three conditions. The selected cross-lagged model must: 1) demonstrate significantly stronger effects compared to a constrained model \cite{kenny1975cross}, as determined by an ANOVA test; 2) have cross-lagged effects that are statistically significant (at least $p<0.05$); and 3) have  the highest absolute mean cross-lagged effect among all the other models. To interpret the values of cross-lagged effect, we employ Cohen's statistical effect size \cite{cohen2013statistical}, where an effect size smaller than 0.2 is considered weak, an effect size between 0.2 and 0.5 is considered moderate, and an effect size between 0.5 and 1.0 is considered strong. The cross-lagged analysis results (Figure \ref{fig:cross-lagged}) indicate that:

\begin{figure}[t]
\centering
\includegraphics[width=0.48\textwidth]{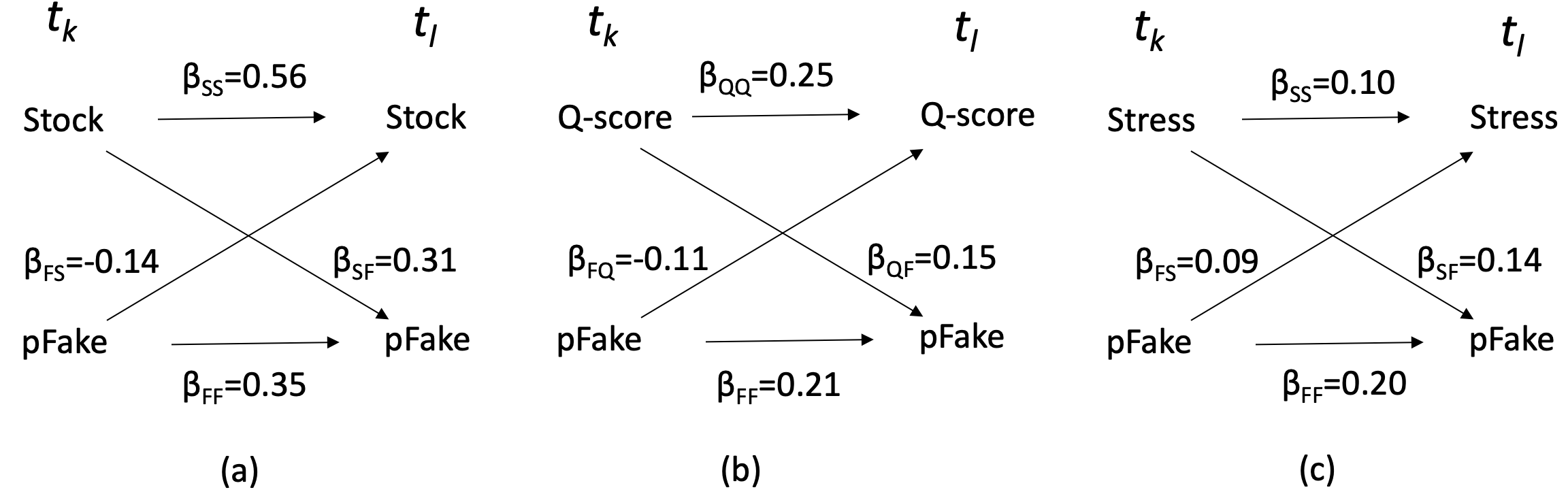}
\caption{\done{Cross-lagged analysis between p\textbf{F}ake and various variables: (a) stock valuation (\textbf{S}tock), (b) external reputation (\textbf{Q}-score) and (c) internal employee stress (\textbf{S}tress).}} 
\label{fig:cross-lagged}
\end{figure}

\begin{description}
    \item \done{\textbf{Stock valuation} (Figure \ref{fig:cross-lagged}a): we found convincing evidence that when a company's stock value goes up, it tends to continue going up in the future. This was shown by a statistical analysis that gave a result of $\beta_{SS} = .56$ with a probability of less than 1\% that it occurred by chance ($p<0.01$). Furthermore, we noticed a moderate tendency for fake news probability (pFake) to reinforce itself ($\beta_{FF} = .35$, $p<0.01$). This implies that when there is a high likelihood of fake news, it is likely to be even higher in the future. In addition, we looked at the relationship between a company's stock value at time $t_k$ and the occurrence of fake news about that company at a later point in time $t_{l}$. The findings showed that there is a moderate, positive connection between the two (standardized $\beta_{SF} = .31$, $p<0.01$). This suggests that when a company's stock value is higher at an earlier point in time, it increases the probability that the company will be the target of fake news at a later point in time. In contrast, we found that exposure to fake news at time  $t_k$ had a weak and negative effect (standardized $\beta_{FS} = -.14$, $p<0.05$) on a company's stock value at a later  time $t_l$. This means that fake news has a detrimental effect (albeit small) on a company's valuation, potentially leading to a decrease in its stock value.}

    \item \done{\textbf{External reputation (Q-score)} (Figure \ref{fig:cross-lagged}b): moderate evidence indicated a positive self-reinforcement of external reputation ($\beta_{QQ} = .25$, $p<0.01$) and pFake ($\beta_{FF} =.21$, $p<0.01$), suggesting that higher levels of external reputation and pFake were associated with subsequent increases in these variables.
    Regarding the cross-lagged effects, 
    we examined the influence of Q-score at $t_k$ on fake news occurrence at $t_l$, where $t_l$ follows $t_k$. Results indicated a weak positive effect ($\beta_{QF} = .15$, $p<0.05$), suggesting that higher Q-scores at an earlier time slightly increased the likelihood of future fake news targeting. Conversely, a weak negative effect ($\beta_{FQ} = -.11$, $p<0.05$) was observed, indicating that fake news exposure at $t_k$ had a small negative impact on subsequent external reputation.}

    \item \done{\textbf{Internal employee stress (Stress)} (Figure \ref{fig:cross-lagged}c):
    there was weak evidence of stress self-reinforcement ($\beta_{SS}=.10$, $p<0.10$), indicating a slight association between higher stress levels and subsequent stress increases. Moderate self-reinforcement of pFake was found ($\beta_{FF}=.20$, $p<0.01$), linking higher pFake levels to increased pFake in the future. Regarding the cross-lagged effects, stress at $t_k$ positively influenced the likelihood of future fake news targeting at $t_l$ with a weak magnitude ($\beta_{SF}=.14$, $p<0.05$). Similarly, fake news exposure at $t_k$ had a weak positive effect on subsequent stress levels at $t_l$($\beta_{FS}=.09$, $p<0.10$).}

\end{description}

\done{To summarize, our cross-lagged analysis revealed that higher past stock valuations increase the future risk of fake news targeting. Weaker effects were observed for external reputation, where higher past reputation makes companies prone to future fake news targeting. Conversely, being targeted by past fake news leads to reputation decline. Regarding internal stress, high past stress levels make companies susceptible to future fake news targeting, and past targeting increases future employee stress. These findings highlight the interplay between stock valuation, reputation, stress, and fake news, indicating risks for targeted companies.}


\section{Discussion}

\noindent \textbf{Limitations.}
Our study has five limitations that should be acknowledged. First, we only attribute fake news to outlets, not individual articles. However, research has shown that publisher-level credibility scores are similar to article-level ones~\cite{xu2021corporate}. Second, our analysis of company perceptions is based on data from Reddit and Glassdoor, so it is unclear if the conclusions hold true on other platforms. Specifically, Reddit is an anonymous platform without the concept of `friends', unlike many other social networks. As such, this platform is missing social pressure, and users are less likely to enter echo chambers, which means that misinformation spreading is not as likely to stem  from homophily \cite{mcpherson2001birds} at the circle-of-friends level as it is on other social networks. 
\done{
Third, we studied the  sentiment of news mentioning a company  and stress mentions in company reviews written by  employees. These aspects are often overlooked. We used state-of-the-art methods to proxy these metrics, and our results show that they are partly predictive of misinformation.  However, our study does not capture all important factors that influence public perception of a company, and might not generalize for other industries or countries.
For example, we did not consider corporate social responsibility \cite{lindgreen2010corporate} or brand identity \cite{balmer2008identity}. These aspects are still subject of ongoing research, and we leave their exploration for future work.}
Fourth, our data did not include comments that were deleted before being collected, so we could not examine if they contained news URLs. In particular, comments that were removed by Automoderator (bots) were unavailable to us, as these comments were removed as soon as they were posted. Nevertheless, the Reddit dataset from ~\url{pushshift.io} remains one of the most comprehensive datasets available~\cite{baumgartner2020pushshift}.
\done{Fifth, our study was conducted on a sample of S\&P 500 companies in the United States. It is possible that our findings may not generalize to other industries, countries, or time periods. Future research is needed to further investigate the generalizability of our findings.}

\hfill \break
\noindent \textbf{Implications.}
Our study has three main \emph{theoretical implications}. 
First, we provide a fine-grained taxonomy of fake news coverage in the corporate world. Second, we identify the factors that contribute to a company's vulnerability to misinformation (e.g., a company's external and internal perceptions). Third, we contribute to understanding the temporal patterns of the misinformation, and how it varies over time. 
Our findings offer an initial step towards the development of a theoretical framework for understanding the complex phenomenon of corporate misinformation.

Our study has four main \emph{practical implications}. 
First, our findings help in identifying which companies are more susceptible to misinformation such as financial performance and reputation, allowing them to take more proactive measures to prevent the spread of false information.
Second, understanding the categories of corporate misinformation that are prevalent, allows companies and stakeholders to better recognize and develop tools and strategies for combating corporate misinformation.
Third, by tracking temporal patterns of misinformation, and detecting ``misinformation shocks'' early on, companies can anticipate and promptly respond to those potential misinformation crises. 
Fourth and final, our results demonstrate the importance of a systematic monitoring system for companies, as well as transparent and direct communication with employees to manage a possible misinformation crisis better and faster \cite{holtom2020tips}.

\hfill \break
\noindent \textbf{Broader perspective, ethics and competing interests.}
\done{Our research originates from a lab within a for-profit, traded company. As such, the authors may benefit financially from the implementation of stronger measures against perpetrators of misinformation. We acknowledge this potential conflict of interest and affirm that the research was conducted with integrity and that the results presented are unbiased and fully reflect all our findings. Our datasets were collected from publicly available sources. We also confirm that our conclusions are based solely on the data and analysis presented, and are not influenced by any financial gain or benefit that may result from the implementation of these measures. We have used rigorous statistical methods to analyze the data and drawn conclusions that are supported by evidence. Furthermore, we will make our data publicly available to allow for replicability.}

\done{Our study was reviewed by the internal company ethical committee while the study protocol, data collection methods, informed consent process, and other aspects of the study  ensured that we met ethical and legal standards. The part of the study that involved human subjects was a crowd-sourcing user study, where participants were paid at least the minimum wage. All our analysis were based on aggregated data without tracking down to individuals.}

\section{Conclusion}
Our study delved into the complex issue of corporate misinformation, specifically focusing on the S\&P 500 companies. Utilizing a combination of large-scale social media and crowd-sourcing data, we uncovered key insights into the nature of fake news targeting these companies. We found that fake news outlets often report on a company's products, politics, and societal issues when spreading misinformation. Furthermore, we established a clear link between a company's external and internal perceptions and its likelihood of being targeted by fake news. Companies that are struggling in the stock market, receiving negative mentions on social media, and whose employees are facing stress are more likely to be targeted by fake news. Additionally, we discovered that fake news about a company does not occur consistently, but rather emerges in specific, concentrated periods. These findings have significant implications for companies and stakeholders looking to combat the spread of misinformation.



\setlength{\bibsep}{0pt plus 0.2ex}
\footnotesize
\bibliography{references}

\begin{thebibliography}{65}
\providecommand{\natexlab}[1]{#1}

\bibitem[{for(2021)}]{forbers21}
 2021.
\newblock GameStop Shares Surge 100\% In A Day, Reddit Group Rejoices.

\bibitem[{Adriani(2019)}]{adriani2019fake}
Adriani, R. 2019.
\newblock Fake news in the corporate world: A rising threat.
\newblock \emph{European Journal of Social Science Education and Research}, 6(1): 92--110.

\bibitem[{Ahmed et~al.(2020)Ahmed, Vidal-Alaball, Downing, Segu{\'\i} et~al.}]{ahmed2020covid}
Ahmed, W.; Vidal-Alaball, J.; Downing, J.; Segu{\'\i}, F.~L.; et~al. 2020.
\newblock COVID-19 and the 5G conspiracy theory: social network analysis of Twitter data.
\newblock \emph{Journal of medical internet research}, 22(5): e19458.

\bibitem[{Akbik et~al.(2019)Akbik, Bergmann, Blythe, Rasul, Schweter, and Vollgraf}]{akbik2019flair}
Akbik, A.; Bergmann, T.; Blythe, D.; Rasul, K.; Schweter, S.; and Vollgraf, R. 2019.
\newblock {FLAIR}: An easy-to-use framework for state-of-the-art {NLP}.
\newblock In \emph{{NAACL} 2019, 2019 Annual Conference of the North American Chapter of the Association for Computational Linguistics (Demonstrations)}, 54--59.

\bibitem[{Al-Rawi and Fakida(2021)}]{al2021methodological}
Al-Rawi, A.; and Fakida, A. 2021.
\newblock The methodological challenges of studying “fake news”.
\newblock \emph{Journalism Practice}, 1--20.

\bibitem[{Alexa.com(2022)}]{alexa}
Alexa.com. 2022.
\newblock Alexa sites.

\bibitem[{Allcott and Gentzkow(2017)}]{allcott2017social}
Allcott, H.; and Gentzkow, M. 2017.
\newblock Social media and fake news in the 2016 election.
\newblock \emph{Journal of economic perspectives}, 31(2): 211--36.

\bibitem[{Allcott, Gentzkow, and Yu(2019)}]{allcott2019trends}
Allcott, H.; Gentzkow, M.; and Yu, C. 2019.
\newblock Trends in the diffusion of misinformation on social media.
\newblock \emph{Research \& Politics}, 6(2): 2053168019848554.

\bibitem[{Balmer(2008)}]{balmer2008identity}
Balmer, J.~M. 2008.
\newblock Identity based views of the corporation: Insights from corporate identity, organisational identity, social identity, visual identity, corporate brand identity and corporate image.
\newblock \emph{European journal of marketing}, 42(9/10): 879--906.

\bibitem[{Balsamo, Bajardi, and Panisson(2019)}]{balsamo2019firsthand}
Balsamo, D.; Bajardi, P.; and Panisson, A. 2019.
\newblock Firsthand opiates abuse on social media: monitoring geospatial patterns of interest through a digital cohort.
\newblock In \emph{The World Wide Web Conference}, 2572--2579.

\bibitem[{Baumgartner et~al.(2020)Baumgartner, Zannettou, Keegan, Squire, and Blackburn}]{baumgartner2020pushshift}
Baumgartner, J.; Zannettou, S.; Keegan, B.; Squire, M.; and Blackburn, J. 2020.
\newblock The pushshift reddit dataset.
\newblock In \emph{Proceedings of the international AAAI conference on web and social media}, volume~14, 830--839.

\bibitem[{Bozarth, Saraf, and Budak(2020)}]{bozarth2020higher}
Bozarth, L.; Saraf, A.; and Budak, C. 2020.
\newblock Higher ground? How groundtruth labeling impacts our understanding of fake news about the 2016 US presidential nominees.
\newblock In \emph{Proceedings of the International AAAI Conference on Web and Social Media}, volume~14, 48--59.

\bibitem[{Bronnenberg, Dub{\'e}, and Sanders(2020)}]{bronnenberg2020consumer}
Bronnenberg, B.~J.; Dub{\'e}, J.-P.; and Sanders, R.~E. 2020.
\newblock Consumer misinformation and the brand premium: A private label blind taste test.
\newblock \emph{Marketing Science}, 39(2): 382--406.

\bibitem[{Camacho and LOZANO(2020)}]{camacho2020disinformation}
Camacho, M.~M.; and LOZANO, A. 2020.
\newblock Disinformation as a corporate risk: the need for a new methodological approach.
\newblock \emph{Vivat Academia}, (155): 111--129.

\bibitem[{Castellani and Berton(2017)}]{castellani2017fake}
Castellani, P.; and Berton, M. 2017.
\newblock Fake news and corporate reputation: What strategies do companies adopt against false information in the media?
\newblock In \emph{Proc. Toulon-Verona Conference" Excellence in Services}.

\bibitem[{Cheng and Chen(2020)}]{cheng2020influence}
Cheng, Y.; and Chen, Z.~F. 2020.
\newblock The influence of presumed fake news influence: Examining public support for corporate corrective response, media literacy interventions, and governmental regulation.
\newblock \emph{Mass Communication and Society}, 23(5): 705--729.

\bibitem[{Chung, Zhang, and Pan(2022)}]{chung2022theory}
Chung, W.; Zhang, Y.; and Pan, J. 2022.
\newblock A Theory-based Deep-Learning Approach to Detecting Disinformation in Financial Social Media.
\newblock \emph{Information Systems Frontiers}, 1--20.

\bibitem[{Cohen(2013)}]{cohen2013statistical}
Cohen, J. 2013.
\newblock \emph{Statistical power analysis for the behavioral sciences}.
\newblock Academic press.

\bibitem[{Cover, Haw, and Thompson(2022)}]{cover2022marginalising}
Cover, R.; Haw, A.; and Thompson, J.~D. 2022.
\newblock Marginalising the Marginalised: Fake News as a Tool of Populist Power.
\newblock In \emph{Fake News in Digital Cultures: Technology, Populism and Digital Misinformation}. Emerald Publishing Limited.

\bibitem[{Di~Domenico and Visentin(2020)}]{di2020fake}
Di~Domenico, G.; and Visentin, M. 2020.
\newblock Fake news or true lies? Reflections about problematic contents in marketing.
\newblock \emph{International Journal of Market Research}, 62(4): 409--417.

\bibitem[{Dot(2022)}]{dailydot}
Dot, D. 2022.
\newblock Fake News Sites list Facebook.

\bibitem[{Eady et~al.(2023)Eady, Paskhalis, Zilinsky, Bonneau, Nagler, and Tucker}]{eady2023exposure}
Eady, G.; Paskhalis, T.; Zilinsky, J.; Bonneau, R.; Nagler, J.; and Tucker, J.~A. 2023.
\newblock Exposure to the Russian Internet Research Agency foreign influence campaign on Twitter in the 2016 US election and its relationship to attitudes and voting behavior.
\newblock \emph{Nature Communications}, 14(1): 62.

\bibitem[{Edge(2021)}]{edge2021scientific}
Edge, L. 2021.
\newblock Scientific misinformation: A perfect storm, missteps, and moving forward.
\newblock \emph{Cell}, 184.

\bibitem[{Fox(2020)}]{fox2020fake}
Fox, J. 2020.
\newblock ‘Fake news’--the perfect storm: historical perspectives.
\newblock \emph{Historical Research}, 93(259): 172--187.

\bibitem[{Freeman and Dart(1993)}]{freeman1993measuring}
Freeman, K.~D.; and Dart, J. 1993.
\newblock Measuring the perceived quality of professional business services.
\newblock \emph{Journal of Professional Services Marketing}, 9(1): 27--47.

\bibitem[{Garimella et~al.(2018)Garimella, De~Francisci~Morales, Gionis, and Mathioudakis}]{garimella2018political}
Garimella, K.; De~Francisci~Morales, G.; Gionis, A.; and Mathioudakis, M. 2018.
\newblock Political discourse on social media: Echo chambers, gatekeepers, and the price of bipartisanship.
\newblock In \emph{Proceedings of the world wide web conference}, 913--922.

\bibitem[{Gillin(2018)}]{gillin2018politifact}
Gillin, J. 2018.
\newblock PolitiFact’s guide to fake news websites and what they peddle. Politifact.

\bibitem[{Graham et~al.(2011)Graham, Nosek, Haidt, Iyer, Koleva, and Ditto}]{graham2011mapping}
Graham, J.; Nosek, B.~A.; Haidt, J.; Iyer, R.; Koleva, S.; and Ditto, P.~H. 2011.
\newblock Mapping the moral domain.
\newblock \emph{Journal of personality and social psychology}, 101(2): 366.

\bibitem[{Holtom, Edmondson, and Niu(2020)}]{holtom2020tips}
Holtom, B.; Edmondson, A.; and Niu, D. 2020.
\newblock Tips for communicating with employees during a crisis.
\newblock \emph{Harvard Business Review}.

\bibitem[{Jahng(2021)}]{jahng2021fake}
Jahng, M.~R. 2021.
\newblock Is fake news the new social media crisis? Examining the public evaluation of crisis management for corporate organizations targeted in fake news.
\newblock \emph{International Journal of Strategic Communication}, 15(1): 18--36.

\bibitem[{Kalsnes(2018)}]{kalsnes2018fake}
Kalsnes, B. 2018.
\newblock Fake news.
\newblock In \emph{Oxford Research Encyclopedia of Communication}.

\bibitem[{Kenny(1975)}]{kenny1975cross}
Kenny, D.~A. 1975.
\newblock Cross-lagged panel correlation: A test for spuriousness.
\newblock \emph{Psychological bulletin}, 82(6): 887.

\bibitem[{K{\i}c{\i}man et~al.(2023)K{\i}c{\i}man, Ness, Sharma, and Tan}]{kiciman2023causal}
K{\i}c{\i}man, E.; Ness, R.; Sharma, A.; and Tan, C. 2023.
\newblock Causal reasoning and large language models: Opening a new frontier for causality.
\newblock \emph{arXiv preprint arXiv:2305.00050}.

\bibitem[{Kogan, Moskowitz, and Niessner(2019)}]{kogan2019fake}
Kogan, S.; Moskowitz, T.~J.; and Niessner, M. 2019.
\newblock Fake news: Evidence from financial markets.
\newblock \emph{SSRN}, 3237763.

\bibitem[{Lazer et~al.(2018)Lazer, Baum, Benkler, Berinsky, Greenhill, Menczer, Metzger, Nyhan, Pennycook, Rothschild et~al.}]{lazer2018science}
Lazer, D.~M.; Baum, M.~A.; Benkler, Y.; Berinsky, A.~J.; Greenhill, K.~M.; Menczer, F.; Metzger, M.~J.; Nyhan, B.; Pennycook, G.; Rothschild, D.; et~al. 2018.
\newblock The science of fake news.
\newblock \emph{Science}, 359(6380): 1094--1096.

\bibitem[{Lin(2016)}]{lin2016new}
Lin, T.~C. 2016.
\newblock The new market manipulation.
\newblock \emph{Emory LJ}, 66: 1253.

\bibitem[{Lindgreen and Swaen(2010)}]{lindgreen2010corporate}
Lindgreen, A.; and Swaen, V. 2010.
\newblock Corporate social responsibility.
\newblock \emph{International journal of management reviews}, 12(1): 1--7.

\bibitem[{MBSFC(2022)}]{mediabias}
MBSFC. 2022.
\newblock Media Bias Fact Check.

\bibitem[{McPherson, Smith-Lovin, and Cook(2001)}]{mcpherson2001birds}
McPherson, M.; Smith-Lovin, L.; and Cook, J.~M. 2001.
\newblock Birds of a feather: Homophily in social networks.
\newblock \emph{Annual review of sociology}, 415--444.

\bibitem[{Melchior and Oliveira(2022)}]{melchior2022health}
Melchior, C.; and Oliveira, M. 2022.
\newblock Health-related fake news on social media platforms: A systematic literature review.
\newblock \emph{new media \& society}, 24(6): 1500--1522.

\bibitem[{Miles(2014)}]{miles2014tolerance}
Miles, J. 2014.
\newblock Tolerance and variance inflation factor.
\newblock \emph{Wiley statsref: statistics reference online}.

\bibitem[{Mills and Robson(2019)}]{mills2019brand}
Mills, A.~J.; and Robson, K. 2019.
\newblock Brand management in the era of fake news: narrative response as a strategy to insulate brand value.
\newblock \emph{Journal of Product \& Brand Management}.

\bibitem[{Moshood, Shittu, and Abidin(2020)}]{moshood2020covid}
Moshood, T.~D.; Shittu, R.~A.; and Abidin, T. 2020.
\newblock Covid-19 and 5G radiation are two parallel lines: A systematic review.
\newblock \emph{Int. J. Innov. Sci. Res. Technol}, 5: 744--751.

\bibitem[{Muzykant et~al.(2021)Muzykant, Muqsith, Pratomo, and Barabash}]{muzykant2021fake}
Muzykant, V.~L.; Muqsith, M.~A.; Pratomo, R.~R.; and Barabash, V. 2021.
\newblock Fake news on COVID-19 in Indonesia.
\newblock In \emph{Pandemic Communication and Resilience}, 363--378.

\bibitem[{Naeem, Bhatti, and Khan(2021)}]{naeem2021exploration}
Naeem, S.~B.; Bhatti, R.; and Khan, A. 2021.
\newblock An exploration of how fake news is taking over social media and putting public health at risk.
\newblock \emph{Health Information \& Libraries Journal}, 38(2): 143--149.

\bibitem[{Ng and Taeihagh(2021)}]{ng2021does}
Ng, L.~H.; and Taeihagh, A. 2021.
\newblock How does fake news spread? Understanding pathways of disinformation spread through APIs.
\newblock \emph{Policy \& Internet}, 13(4): 560--585.

\bibitem[{Park et~al.(2020)Park, Montecchi, Plangger, Pitt et~al.}]{park2020understanding}
Park, A.; Montecchi, M.; Plangger, K.; Pitt, L.; et~al. 2020.
\newblock Understanding fake news: a bibliographic perspective.
\newblock \emph{Defence Strategic Communications}, 8(Spring 2020): 141--172.

\bibitem[{Pennycook and Rand(2021)}]{pennycook2021psychology}
Pennycook, G.; and Rand, D.~G. 2021.
\newblock The psychology of fake news.
\newblock \emph{Trends in cognitive sciences}, 25(5): 388--402.

\bibitem[{Pickard(2019)}]{pickard2019misinformation}
Pickard, V. 2019.
\newblock The Misinformation Society.
\newblock In \emph{Antidemocracy in America}, 39--48. Columbia University Press.

\bibitem[{PWC(2022)}]{pwc}
PWC. 2022.
\newblock Corporate Sector Disinformation.

\bibitem[{Scepanovic et~al.(2020)Scepanovic, Martin-Lopez, Quercia, and Baykaner}]{scepanovic2020extracting}
Scepanovic, S.; Martin-Lopez, E.; Quercia, D.; and Baykaner, K. 2020.
\newblock Extracting medical entities from social media.
\newblock In \emph{Proceedings of the ACM Conference on Health, Inference, and Learning}, 170--181.

\bibitem[{Serazio(2021)}]{serazio2021other}
Serazio, M. 2021.
\newblock The other ‘fake’news: Professional ideals and objectivity ambitions in brand journalism.
\newblock \emph{Journalism}, 22(6): 1340--1356.

\bibitem[{Starbird(2021)}]{starbird2021online}
Starbird, K. 2021.
\newblock Online Rumors, Misinformation and Disinformation: The Perfect Storm of COVID-19 and Election 2020.

\bibitem[{Swire-Thompson and Lazer(2019)}]{swire2019public}
Swire-Thompson, B.; and Lazer, D. 2019.
\newblock Public health and online misinformation: challenges and recommendations.
\newblock \emph{Annual review of public health}, 41: 433--451.

\bibitem[{Tandoc~Jr(2019)}]{tandoc2019facts}
Tandoc~Jr, E.~C. 2019.
\newblock The facts of fake news: A research review.
\newblock \emph{Sociology Compass}, 13(9): e12724.

\bibitem[{Vargo, Guo, and Amazeen(2018)}]{vargo2018agenda}
Vargo, C.~J.; Guo, L.; and Amazeen, M.~A. 2018.
\newblock The agenda-setting power of fake news: A big data analysis of the online media landscape from 2014 to 2016.
\newblock \emph{New media \& society}, 20(5): 2028--2049.

\bibitem[{Verma et~al.(2022)Verma, Bhardwaj, Aledavood, De~Choudhury, and Kumar}]{verma2022examining}
Verma, G.; Bhardwaj, A.; Aledavood, T.; De~Choudhury, M.; and Kumar, S. 2022.
\newblock Examining the impact of sharing COVID-19 misinformation online on mental health.
\newblock \emph{Scientific Reports}, 12(1): 1--9.

\bibitem[{Vosoughi, Roy, and Aral(2018)}]{vosoughi2018spread}
Vosoughi, S.; Roy, D.; and Aral, S. 2018.
\newblock The spread of true and false news online.
\newblock \emph{science}, 359(6380): 1146--1151.

\bibitem[{Waisbord(2018)}]{waisbord2018truth}
Waisbord, S. 2018.
\newblock Truth is what happens to news: On journalism, fake news, and post-truth.
\newblock \emph{Journalism studies}, 19(13): 1866--1878.

\bibitem[{Westerlund(2019)}]{westerlund2019emergence}
Westerlund, M. 2019.
\newblock The emergence of deepfake technology: A review.
\newblock \emph{Technology Innovation Management Review}, 9(11).

\bibitem[{Wikipedia(2022)}]{wikisp}
Wikipedia. 2022.
\newblock S\&P 500 Companies.

\bibitem[{Xu(2021)}]{xu2021corporate}
Xu, R. 2021.
\newblock \emph{Corporate Fake News on Social Media}.
\newblock Ph.D. thesis, University of Miami.

\bibitem[{{Yahoo Finance portal}(2022)}]{yahoo}
{Yahoo Finance portal}. 2022.
\newblock S\&P 500 Stock Market Data.

\bibitem[{Zhou and Zafarani(2020)}]{zhou2020survey}
Zhou, X.; and Zafarani, R. 2020.
\newblock A survey of fake news: Fundamental theories, detection methods, and opportunities.
\newblock \emph{ACM Computing Surveys (CSUR)}, 53(5): 1--40.

\bibitem[{Zimdars(2016)}]{zimdars2016my}
Zimdars, M. 2016.
\newblock My “fake news list” went viral. But made-up stories are only part of the problem.
\newblock \emph{Washington Post}.

\end{thebibliography}

\normalsize
\appendix

\section{Appendix}

Our paper has five main findings\footnote{\url{https://social-dynamics.net/fake-news/}}:
\begin{enumerate}
\item Companies that suffer severely from fake news are also likely to be covered by reputable news publishers. 
\item Companies susceptible to fake news receive less attention from the general public on social media.
\item Being targeted by fake news leads to a higher likelihood of public comments with negative sentiments on social media.
\item A company targeted by fake news is likely to experience a lower stock market growth.
\item In companies targeted by fake news, employees are more likely to express stress indicators in their reviews. 
\end{enumerate}

In recent advancements, the efficacy of Large Language Models (LLMs) in elucidating causal relationships has emerged as a potential application \cite{kiciman2023causal}. 
Building upon this work \cite{kiciman2023causal}, we embark on an exploration of causal analysis leveraging the capabilities of GPT4, a state-of-the-art language model\footnote{\url{https://openai.com/blog/gpt-4-api-general-availability}}. Our primary objective revolves around validating our five findings. To achieve this,  we adopt prompt strategies from \cite{kiciman2023causal} aimed at validating each causal claim by evaluating its normality or abnormality. A typical prompt includes:

\begin{description}
\item \emph{System}: You are a helpful assistant for causal reasoning. Specifically, you are an advanced expert specializing in causal reasoning within the realm of Corporate Misinformation research using public data, such as stock price, published news, and social media responses. Equipped with a strong knowledge base and experience in financial, social and environmental determinants of misinformation in the company world, you are proficient in identifying, assessing, and interpreting the complex interactions among variables. Your primary responsibility involves validating known or hypothesized causal relationships between a diverse array of variables and their respective outcomes.

\item \emph{User:}
You will receive a claim on the cause-and-effect relationship between variables. Based on all the information, given a causal event that causes an outcome event at a later time point, the causal event is ``abnormal'' if the causal event’s occurrence was unexpected, unlikely, surprising, rare, or improbable. The opposite of ``abnormal'' is ``normal''. A causal event is `normal' if its occurrence was expected, likely, probable, and not surprising. The normality of a causal event is independent of whether the outcome event was bad or and whether the outcome event was normal. Please read the following statement about a causal event. Provide your step-by step rationale as to why the causal event is abnormal or normal. Then, return the word ``normal'' if you think the causal event is normal or ``abnormal'' if you think the causal event is abnormal. 

\item \emph{Claim}: \emph{Companies that suffer severely from fake news are also likely to be covered by reputable news publishers. }

\end{description}

The step-by-step answer of GPT-4 to that typical prompt unfolds as follows: ``The claim suggests a causal relationship between severe suffering from fake news by companies and increased coverage by reputable news publishers. Let's break this down step by step:

\begin{description}
\item \emph{Companies suffering severely from fake news}: This implies a situation where a company is significantly affected by false information spread about it. 

\item \emph{Likelihood of coverage by reputable news publishers}: The claim implies that when a company faces severe fake news attacks, reputable news outlets are more likely to cover the situation.

\item \emph{Rationale for Abnormality or Normality}:
\begin{itemize}
\item \emph{Abnormal}: This relationship might be considered abnormal because, traditionally, reputable news publishers tend to verify and fact-check information rigorously before reporting. Therefore, if a company is suffering from fake news, the expectation might be that reputable news sources would be cautious about covering unverified or misleading information, potentially leading to less coverage rather than more.

\item \emph{Normal}: On the other hand, in today's media landscape, there's a growing emphasis on reporting on significant events, even if they are controversial or involve misinformation. Companies facing severe fake news might attract attention due to the sensational nature of the situation, prompting reputable news outlets to cover it to clarify the facts and counteract the false information.
\end{itemize}

\item Considering the dynamics of sensationalism in media and the role of reputable news sources in countering fake news, the relationship might indeed be normal. While it might seem unexpected for reputable sources to cover such companies more, the desire to address and rectify false information could drive increased coverage. \emph{Therefore, the claim is \emph{normal} in this scenario.}

\end{description}

\begin{table*}[t]
\small
\centering
\begin{tabularx}{0.95\textwidth}{|X|X|}
\hline
 {\bf Statement} & {\bf Conclusion }
 \\ \hline
 Companies that suffer severely from fake news are also likely to be covered more by reputable news publishers. &  Considering the dynamics of sensationalism in media and the role of reputable news sources in countering fake news, the relationship might indeed be normal. While it might seem unexpected for reputable sources to cover such companies more, the desire to address and rectify false information could drive increased coverage. \emph{Therefore, the claim is {\bf \emph{normal}} in this scenario.}
 \\ \hline
 Companies susceptible to fake news receive less attention from the general public on social media. &  Considering the dynamics of how fake news spreads on social media and the potential hesitancy to engage with content that might be false, the relationship between companies being subject to fake news and the general public talking less about them on social media might be considered normal. The decrease in discussion could be a result of cautious behavior among the public to avoid amplifying false information. \emph{Therefore, the claim is {\bf \emph{normal}} in this scenario.}  \\ \hline
 Being targeted by fake news leads to a higher likelihood of public comments with negative sentiments on social media, contributing to the decline in the reputation of affected companies. & Considering the typical response patterns to fake news and its impact on reputation, the relationship between being targeted by fake news, experiencing reputation decline, and receiving negative comments on social media is more likely to be normal. The emotional response triggered by fake news and its effect on public sentiment could drive increased negative comments on social media. \emph{Therefore, the claim is {\bf \emph{normal}} in this scenario.}  \\ \hline
A company targeted by fake news is likely to experience a lower stock market growth.  & Considering the complexities of the stock market and the multitude of factors influencing stock prices, the relationship between being likely to be targeted by fake news and maintaining lower stock growth might be normal. While other market dynamics are at play, the impact of negative news or misinformation on stock performance could lead to lower growth. \emph{Therefore, the claim is {\bf \emph{normal}} in this scenario.}  \\ \hline
In companies targeted by fake news, employees are more likely to express stress indicators in their reviews.  & Considering the potential impact of negative external events on employee morale and the prevalence of stress-related discussions in workplace environments, the relationship between companies susceptible to fake news targeting and considerable stress mentions by their employees might be considered normal. The spread of false information about the company could indeed impact the internal environment and employee well-being. \emph{Therefore, the claim is {\bf \emph{normal}} in this scenario.}  \\
\hline
 Employees express stress indicators when reviewing their workplaces are more likely to be targeted by fake news. & Considering the link between employee stress indicators and potential targeting by fake news, it might be considered abnormal as the direct causal relationship between these variables could be unexpected or unlikely. The connection might involve other complex factors beyond just stress indicators.
\emph{Therefore, the causal event is {\bf \emph{abnormal}} in this scenario.} 
\\ \hline
\end{tabularx}
\caption{Validation of our five main findings and one unsubstantiated claim by prompting LLM GPT-4 for causality as per \cite{kiciman2023causal}. Causal relationships are considered validated if found to be normal.}
\label{tab:causal_validation}
\end{table*}

\vspace{0.2in}

The resulting assessments of the (ab)normality of our five claims (Table \ref{tab:causal_validation}) suggest that all five of our main findings are supported. However, the LLM does not always return 'normal' as an answer. To illustrate this, we also considered statements for which our causal analysis did not have strong evidence to be conclusive. One such statement is reported in the last row of Table \ref{tab:causal_validation} and suggests that companies in which employees express stress indicators when reviewing their workplaces are more likely to be targeted by fake news. Indeed, the LLM found this claim to be abnormal.

\end{document}